\PassOptionsToPackage{table}{xcolor} %
\documentclass[sigconf]{acmart}
\usepackage{tikz}
\usepackage{xcolor} %
\usetikzlibrary{calc}

\usepackage{framed}
\usepackage{acmart-taps}

\usepackage{caption}

\usepackage[final,nocomment,commandnameprefix=ifneeded]{changes}
\usepackage{framed}   %
\usepackage{graphicx} %
\makeatletter

\usepackage{ifthen}

\newboolean{showrevisions}
\setboolean{showrevisions}{false}  %

\definecolor{revadd}{RGB}{0,64,200}   %
\definecolor{revdel}{RGB}{200,0,0}    %
\definecolor{revnote}{RGB}{0,128,0}   %

\newcommand{\rev}[1]{%
  \ifthenelse{\boolean{showrevisions}}%
    {\begingroup\color{revadd}#1\endgroup}%
    {#1}%
}

\newcommand{\revdel}[1]{%
  \ifthenelse{\boolean{showrevisions}}%
    {\begingroup\color{revdel}\sout{#1}\endgroup}%
    {}%
}

\definecolor{muisecondary}{HTML}{9500ae}
\definecolor{muiprimary}{HTML}{2196f3}
\definecolor{mygrey}{HTML}{808080}

\newcommand{\SideLabelSep}{2pt}   %
\newcommand{\SideTextSep}{3pt}    %
\newcommand{\SideRuleWidth}{2pt} 

\newenvironment{SideBarBlock}[2]{%
  \def\FrameCommand{%
    \hspace*{-\SideTextSep}\hspace*{-\SideRuleWidth}%
    \makeatletter
    \if@firstcolumn
      \llap{%
        \rotatebox{90}{\textcolor{#1}{\sffamily\bfseries\footnotesize #2}}%
        \hspace{\SideLabelSep}%
      }%
      {\color{#1}\vrule width \SideRuleWidth}%
      \hspace{\SideTextSep}%
    \else
      \hspace{\hsize}%
      \hspace{\SideTextSep}%
      {\color{#1}\vrule width \SideRuleWidth}%
      \rlap{%
        \hspace{\SideLabelSep}%
        \rotatebox{90}{\textcolor{#1}{\sffamily\bfseries\footnotesize #2}}%
      }%
      \hspace{-\hsize}\hspace{-\SideTextSep}\hspace{-\SideRuleWidth}%
    \fi
    \makeatother
  }%
  \MakeFramed{\advance\hsize-\width \FrameRestore}%
  \setlength{\parindent}{0pt}%
}{%
  \endMakeFramed%
}

\aptLtoXcmd{\newenvironment{expert}
  {\begin{leftbar}%
  \textcolor{muisecondary}{\textbf{EXPERT}}\newline\ }
  {\end{leftbar}}}{
  \newenvironment{expert}
  {\begin{SideBarBlock}{muisecondary}{EXPERT}}
  {\end{SideBarBlock}}
  }

\aptLtoXcmd{\newenvironment{overall}
  {\begin{leftbar}%
  \textcolor{mygrey}{\textbf{OVERALL}}\newline}
  {\end{leftbar}}}{\newenvironment{overall}
  {\begin{SideBarBlock}{mygrey}{OVERALL}}
  {\end{SideBarBlock}}}

\aptLtoXcmd{\newenvironment{cs}
  {\begin{leftbar}%
  \textcolor{muiprimary}{\textbf{COMPARATIVE~STUDY}}\newline}
  {\end{leftbar}}}{\newenvironment{cs}
  {\begin{SideBarBlock}{muiprimary}{\shortstack{COMPARATIVE\\[-2pt]STUDY}}}
  {\end{SideBarBlock}}}

\aptLtoXcmd{\newenvironment{lw}
  {\begin{leftbar}}%
  {\end{leftbar}}}{\newenvironment{lw}
  {\begin{SideBarBlock}{white}{}}
  {\end{SideBarBlock}}}

\AtBeginDocument{%
  }

\copyrightyear{2026}
\acmYear{2026}
\setcopyright{cc}
\setcctype{by}
\acmConference[CHI '26]{Proceedings of the 2026 CHI Conference on Human Factors in Computing Systems}{April 13--17, 2026}{Barcelona, Spain}
\acmBooktitle{Proceedings of the 2026 CHI Conference on Human Factors in Computing Systems (CHI '26), April 13--17, 2026, Barcelona, Spain}
\acmPrice{}
\acmDOI{10.1145/3772318.3790793}
\acmISBN{979-8-4007-2278-3/2026/04}

\usepackage{booktabs}

\begin{document}

\title[\emph{GroundLink}]{\emph{GroundLink}: Exploring How Contextual Meeting Snippets\\ Can Close Common Ground Gaps in Editing 3D Scenes\\ for Virtual Production}

\author{Gun Woo (Warren) Park}
\orcid{0009-0001-6187-7688}
\affiliation{\institution{Autodesk Research}
\city{Toronto}
\state{Ontario}
\country{Canada}}
\affiliation{
\department{Department of Computer Science}\institution{University of Toronto}
\city{Toronto}
\state{Ontario}
\country{Canada}}
\email{warren@dgp.toronto.edu}

\author{Frederik Brudy}
\orcid{0000-0002-3868-0967}
\affiliation{\institution{Autodesk Research}
\city{Toronto}
\state{Ontario}
\country{Canada}}
\email{frederik.brudy@autodesk.com}

\author{George Fitzmaurice}
\orcid{0000-0002-2834-7757}
\affiliation{\institution{Autodesk Research}
\city{Toronto}
\state{Ontario}
\country{Canada}}
\email{george.fitzmaurice@autodesk.com}

\author{Fraser Anderson}
\orcid{0000-0003-3486-8943}
\affiliation{\institution{Autodesk Research}
\city{Toronto}
\state{Ontario}
\country{Canada}}
\email{fraser.anderson@autodesk.com}

\renewcommand{\shortauthors}{Park et al.}

\begin{abstract}
Virtual Production (VP) professionals often face challenges accessing tacit knowledge and creative intent, which are important in forming common ground with collaborators and in contributing more effectively and efficiently to the team. From our formative study (N=23) with a follow-up interview (N=6), we identified the significance and prevalence of this challenge. To help professionals access knowledge, we present \emph{GroundLink}, a Unity add-on that surfaces meeting-derived knowledge directly in the editor to support establishing common ground. It features a meeting knowledge dashboard for capturing and reviewing decisions and comments, constraint-aware feedforward that proactively informs the editor environment, and cross-modal synchronization that provides referential links between the dashboard and the editor. A comparative study (N=12) suggested that \emph{GroundLink} help users build common ground with their team while improving perceived confidence and ease of editing the 3D scene. An expert evaluation with VP professionals (N=5) indicated strong potential for \emph{GroundLink} in real-world workflows.

\end{abstract}


\begin{teaserfigure}
\centering
  \includegraphics[width=0.75\textwidth]{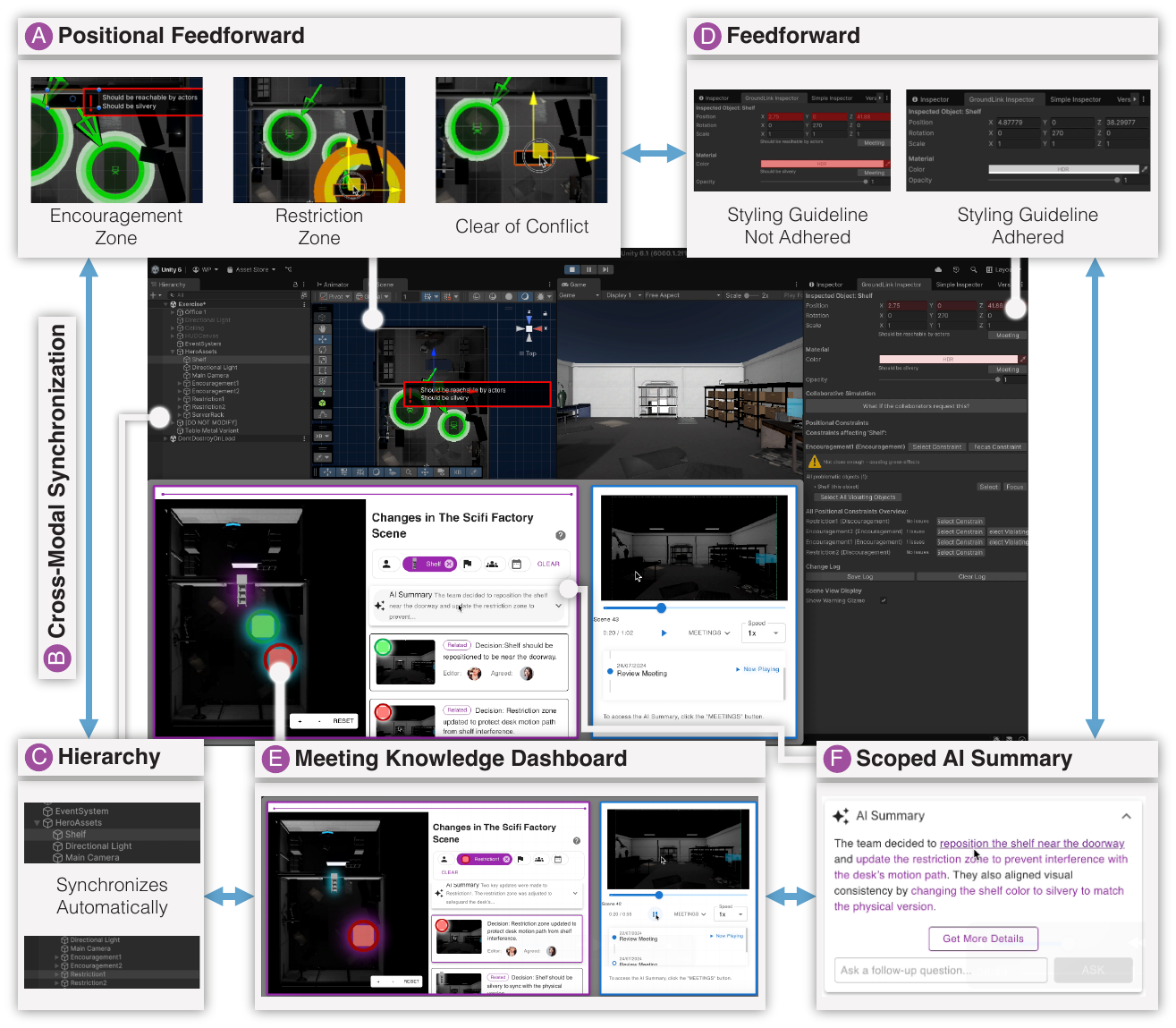}
  \caption{\emph{GroundLink} surfaces decisions and rationale behind edits made in past meetings in virtual production projects, right within the Unity editor. (A) Positional feedforward: When the placement of an element does not follow a previous decision, it shows a green circle with an arrow to encourage placement in an encouragement zone, or a yellow/orange/red circle to encourage moving away from the restriction zone. If the placement follows the decision, all zones remain transparent. (B) Importantly, all the components in \emph{GroundLink} synchronize, whether it is verbal (meeting recordings, summaries) or non-verbal content (3D editor). (C) The Hierarchy view in Unity also syncs, as does the Inspector (D). Just like positional feedforward, the Inspector blinks the field in red if the properties do not follow the styling guideline or decisions. (E) The meeting knowledge dashboard displays all the meeting decisions and summaries in an easily navigable format. (F) It also features an AI summary, which users can ask more scoped questions about objects and collaborators among all decisions and comments.
  }
  \Description{The figure shows the Unity desktop editor with the GroundLink add-on enabled. Six components are displayed. A shows a positional feedforward, where the placement of an object may be encouraged, restricted, or left unaffected by prior decisions. B shows a line connecting all six components, indicating that they are linked in a cross-modal way. C shows the Unity hierarchy view, which is also synchronized automatically when other components are interacted with. E shows the meeting knowledge dashboard, where users can access past meeting summaries and play meeting recordings. Within it, there is a scoped AI summary (F) that allows users to obtain a focused summary of the decisions and comments from remote meetings.}
  \label{fig:teaser}
\end{teaserfigure}

\maketitle

\section{Introduction}
Virtual Production (VP), a rapidly growing branch of filmmaking, requires continuous, iterative coordination between artists, technicians, and supervisors throughout the entire filmmaking process, promising creative freedom and cost savings~\cite{zwerman_ves_2023,kadner_noah_virtual_2021}. This coordination relies heavily on frequent meetings where teams make important decisions about creative intent, technical constraints, asset specifications, and workflow priorities. These meetings serve as the primary mechanism for establishing shared understanding and aligning diverse expertise across departments, from lighting designers and camera operators to VFX artists and technical directors.

However, the decisions made in these meetings often become disconnected from the actual production work. For example, during intensive creation phases, team members could focus on immediate technical tasks and may forget or misinterpret earlier agreements. The increasing trend toward remote and distributed VP workflows~\cite{swords_emergence_2024,swords_it_2024,barnett_beyond_2024} can amplify this challenge. While remote meetings enable global collaboration with top talent, they often result in less effective knowledge sharing compared to in-person interactions~\cite{niemantsverdriet_recurring_2017}. 

The challenge intensifies when a \emph{newcomer}\footnote{Can be an experienced professional.} joins an ongoing project. Newcomers inherit work shaped by dozens of prior meetings but lack efficient access to the rationale behind existing decisions. 
This creates a significant knowledge gap: important decisions and constraints discussed in meetings remain buried in recordings while team members work in 3D environments where those decisions should be~applied.

Literature suggests that knowledge workers, in general (which the VP professionals are), rarely start a project with sufficient access to materials, expertise, or experience~\cite{nonaka_knowledge-creating_1998,garicano_organization_2006,argote_knowledge_2000}, and that successful teams overcome these imbalances by establishing \emph{common ground}: the shared understandings, tacit know-how, preferences, and constraints that allow members to communicate and coordinate smoothly~\cite{clark_herbert_h_communities_1996,olson_judith_s_common_2014,barnlund_transactional_1970}. However, it remains unclear whether VP professionals actually experience common ground issues in practice, and if so, what specific types of meeting-derived knowledge are most important for establishing effective collaboration.

We address these questions by investigating how to surface meeting-derived knowledge directly within the 3D content creation environment. Specifically, we ask:
\begin{itemize} 
\item \textbf{RQ1:} What kinds of meeting-derived knowledge do virtual production teams need when working on 3D scenes? 
\item \textbf{RQ2:} How does surfacing meeting-derived knowledge support a team member in virtual production projects form common ground with existing team members? 
\item \textbf{RQ3:} How does acquiring knowledge from past meeting recordings affect team members' perceived confidence and task difficulty? 
\end{itemize}

To answer \textbf{RQ1}, we first conducted a formative study, with experienced VP professionals (N=23) and follow-up interview (N=6). This study confirmed that VP professionals do indeed experience hindrance in forming common grounds, particularly when newcomers join projects, and revealed the specific types of tacit knowledge and collaborative intent that are typically shared through meetings but become difficult to access during production work.

Guided by these empirical findings, we designed and developed \emph{GroundLink}, a Unity add-on that augments 3D scenes with contextual meeting snippets. \emph{GroundLink} works with existing, indexed video meeting recordings and surfaces relevant meeting snippets as users navigate and interact with 3D scene elements, creating bidirectional connections between the scene and meeting content. As users edit the scene, \emph{GroundLink} provides contextual cues during editing, surfacing potential conflicts with earlier decisions while preserving user agency for creative judgment. It also allows users to generate summaries of past changes to help them understand constraints and collaboration~needs.

To answer \textbf{RQ2} and \textbf{RQ3}, we evaluated \emph{GroundLink} through two complementary studies. First, we conducted an exploratory, comparative within-subjects study (N=12) with participants who have 3D editing experience but no VP domain expertise. Participants edited two unfamiliar Unity scenes under time pressure, using both a powerful baseline (Microsoft Copilot) and \emph{GroundLink}. This design allowed us to isolate the effects of our approach from confounds related to VP skill level or project familiarity. Results show that participants reported more extensive perceived common ground with the prior team and greater confidence in the quality of their results, when using \emph{GroundLink}. To confirm domain-specific validity, we also conducted an expert evaluation (N=5) with VP professionals, who confirmed that \emph{GroundLink} has potential to address real challenges in their workflows and would be valuable for their collaborative~practices.

This paper contributes: 
\begin{itemize} 
\item \emph{GroundLink}, a Unity add-on that helps distributed collaborators build common ground by surfacing meeting-derived knowledge within 3D editing workflows. 
\item Empirical findings from a formative study revealing specific knowledge gaps in virtual production collaboration and confirming that issues in forming common ground identified in literature do manifest in practice. 
\item A comparative user study demonstrating the \added{perceived} effectiveness of in-situ knowledge surfacing for newcomer onboarding in collaborative 3D environments, as well as an expert evaluation confirming the potential of the approach in real-world workflows.
\item Design implications for integrating meeting-derived knowledge into production tools, applicable to virtual production, and other collaborative 3D workflows. 
\end{itemize}
\section{Background}

This work addresses the challenges in virtual production. It builds on prior research in collaboration, with particular emphasis on remote work and relevant designs (feedforward and in-situ presentation of~information).

\subsection{Virtual Production and Collaboration} 
\label{sect:b-vpc} 
Virtual production (VP) is a fast-growing filmmaking technique that blends live-action footage with computer-generated (CG) content in real time~\cite{zwerman_ves_2023,kadner_noah_virtual_2021}. Using technologies like LED volume stages and motion capture systems, filmmakers can create high-fidelity virtual environments and props that would be difficult or expensive to build physically. These assets can be viewed and modified interactively on set, enabling rapid iteration and alignment with creative goals~\cite{swords_it_2024,swords_emergence_2024}. 
VP enables continuous, iterative coordination among artists, technicians, and supervisors throughout the entire process~\cite{kadner_noah_virtual_2021,swords_it_2024}. With VP, professionals can more easily hire top talent globally by using remote work arrangements. The industry is seeing a growing trend of geographically distributed teams, where collaborators work remotely across countries and time zones~\cite{kavakli_virtual_2022}. As a result, despite its relatively recent emergence, VP is increasingly being adopted~\cite{li2022development}.

However, the literature suggests that VP professionals may experience issues with establishing common ground. The film industry, in general, relies heavily on project-based hiring~\cite{blair2001working,defillippi1998paradox,jones1996careers,handy2017systems}, i.e., new project brings together new collaborators who may have never worked together before. Establishing common ground in this setting involves tacit knowledge about naming conventions, artistic preferences, technical constraints, and shared expectations. When common ground is not well established, it is often difficult for collaborators to communicate effectively~\cite{olson_judith_s_common_2014,barnlund_transactional_1970,olson2000distance}. This problem can be amplified in VP, which is at its infancy. 

Remote collaboration adds another layer of complexity. VP allows for hiring top talent from around the world, but this necessitates remote collaboration, which is a growing trend~\cite{swords_emergence_2024,swords_it_2024,barnett_beyond_2024}. Although real-time tools like video conferencing have become standard, they are not always sufficient for communicating nuanced design decisions or creative intent~\cite{daft1986organizational}. 

As prior work notes, remote meetings can often hinder the establishment of strong common ground~\cite{olson2000distance}, especially in recurring meeting settings~\cite{niemantsverdriet_recurring_2017}. Moreover, remote meetings are often viewed as disruptive or time-consuming~\cite{microsoft2023willai,lucid_howmanymeetings_2022,doodle__meetingsgeneral_2019}, potentially leading to fewer meetings and greater divergence in mental models than needed\footnote{It is easy to think that there are too many meetings in remote work~\cite{bailenson2021nonverbal}, but the term `meetings' here includes spontaneous interactions, such as water cooler discussions. When considering this broader definition, we can recognize a reduction in the frequency, quality, and duration of interactions with collaborators~\cite{niemantsverdriet_recurring_2017}.}. However, remote collaboration is a growing standard in VP~\cite{swords_emergence_2024,swords_it_2024,barnett_beyond_2024}, and we therefore consider mechanisms to extract maximum insight from this scarcer source of data.

Given these challenges, we question whether existing collaboration tools are sufficient to support richer collaboration in the growing, remote virtual production field. Our formative study aims to determine whether common ground formation problems manifest in practice for virtual production professionals, closing a research gap where past work has only speculated about such issues. 

\subsection{Common Ground in (Remote) Work} 

\subsubsection{Components of Common Ground in Virtual Production} Common ground is important for effective collaboration, especially in distributed or high-stakes work environments~\cite{olson_judith_s_common_2014,olson2000distance,barnlund_transactional_1970,convertino2008articulating,Clark}. While cognition is often conceptualized as an individual process, it frequently involves knowledge distributed across a team~\cite{rogers_distributed_1994}. It encompasses shared understandings of terminology, goals, constraints, and tacit knowledge that enable teams to communicate and coordinate smoothly. From a cognitive perspective, two complementary theoretical models help explain how common ground forms and functions: Shared Mental Models (SMM) and Transactive Memory Systems (TMS).

SMM emphasizes that collaborators benefit from overlapping knowledge of tasks, processes, and team dynamics. Effective teamwork stems from shared understanding, i.e., everyone being ``on the same page.'' Practices such as planning sessions, reflexive check-ins, and team interaction training have been shown to improve coordination and performance in ad hoc and virtual teams~\cite{cannon1993shared,park_retrospector_2023}.

In contrast, TMS describes common ground in terms of distributed cognition: instead of everyone knowing the same things, team members maintain an awareness of who knows what~\cite{wegner1987transactive}. This model is especially valuable when collaborators have deep specializations, and when sharing all relevant knowledge across the team would be inefficient. TMS assumes that effective coordination relies on knowing whom to consult, rather than on uniform expertise.

Virtual production demands a balance of both paradigms, as different collaboration problems emphasize each one differently. This balance informs the direction of support that our work aims to provide. Professionals bring deep, domain-specific skills (e.g., lighting, rigging, camera tracking), highlighting the importance of TMS. Simultaneously, virtual production involves frequent, close, and cross-functional coordination to realize a shared creative intent~\cite{kadner_noah_virtual_2021}, underscoring the importance of SMM. These theories suggest that a support system for common ground formation in virtual production collaboration should both provide a shared repository of knowledge and offer role-based references to help users identify experts for specific aspects of production.

\subsubsection{Previous Work on Supporting Common Ground Formation} 
Prior systems have focused on capturing and presenting knowledge across different media and contexts to support common ground formation. Common ground can be captured passively or proactively. Passive approaches mine knowledge from corporate documentation, collaborative tools, or archived communication logs~\cite{cortinas2024through,meixner2017chat2doc,wang2022group,brennan2018automatic}. Proactive methods, such as think-aloud computing, encourage users to verbalize their thought processes as they work, making that knowledge more accessible to others or to intelligent systems~\cite{krosnick_think-aloud_2021}. Given that meetings are abundant in virtual production~\cite{kadner_noah_virtual_2021} and that much of the constituent knowledge for common ground is potentially verbalized there, we focus our contribution on helping professionals consume this rich source of knowledge effectively.

Past work has presented this knowledge from meeting recordings using enhanced representations. One approach represents meetings in more digestible formats. Beyond traditional text-based summaries~\cite{li2019keep,liu2008correlation,waibel1998meeting,rennard2023abstractive,murray2005extractive}, MeetMap uses a large language model to generate a dynamic visual map of conversation topics based on issue-based information systems~\cite{kunz1970issues}, for in-meeting awareness and post-meeting review~\cite{chen2025meetmap}. Another approach provides more efficient replay mechanisms for video~\cite{junuzovic2011did,inkpen2010air}, audio~\cite{tucker2010catchup}, or team-shared annotations~\cite{nathan2012case}. These replays can also be spatialized using immersive mixed reality, showing what happened in a space while users were engaged in another task~\cite{fender_causality-preserving_2022,cho_realityreplay_2023}.

Instead of consuming knowledge on a separate occasion, it can be embedded in conversations to help users consume meeting knowledge during the meeting itself. In cases where users miss live meetings, proxies such as AI avatars~\cite{leong2024dittos} or pre-recorded videos~\cite{tang2012time} simulate participation. Knowledge presentation can also be proactive: chatbots can surface relevant information~\cite{birnholtz2005grounding,long2025feedquac,shin2022chatbots}, systems can trigger alerts when misunderstandings arise~\cite{kim_improving_2016}, and AR overlays can provide contextual guidance~\cite{xu2024can}.

In existing systems, the consumption of constituent knowledge for common ground formation remains separate from actual work. A missed opportunity lies in embedding common ground formation into the work itself, where the act of working could simultaneously help users acquire this knowledge. This explores a direction beyond embedding knowledge during meetings, focusing instead on asynchronous work.
This opportunity is particularly significant for remote virtual production professionals, where relying on communication alone requires additional effort to proactively engage with shared resources~\cite{niemantsverdriet_recurring_2017}, and interpreting knowledge in practice is difficult given emerging terminologies and workflows~\cite{kadner_noah_virtual_2021,zwerman_ves_2023}.

\subsection{Surfacing Hidden Knowledge} 
Much of the knowledge necessary for forming common ground is tacit or hidden, residing in prior decisions, implicit intent, or shared understanding that is often not explicitly documented~\cite{olson_judith_s_common_2014}. In co-located teams, such knowledge is shared informally through spontaneous conversations or observations. In remote settings, these opportunities are reduced~\cite{niemantsverdriet_recurring_2017}. Our goal is to help virtual production professionals obtain this constituent knowledge in situ, building on two relevant concepts: feedforward and the in-situ presentation of information.

\subsubsection{Feedforward} 
Feedforward refers to interface mechanisms that preview the outcome of an interaction before the user acts, helping them anticipate consequences~\cite{vermeulen_crossing_2013,norman1988psychology,djajadiningrat2002but}. Originally used in gesture guidance~\cite{bau2008octopocus,fennedy2021octopocus,park2024jollygesture,khurana2024just}, VR interaction guidance~\cite{muresan2023using,artizzu2024virgilites}, and GUI interactions~\cite{coppers2019fortunettes,terry2002side}, it has recently expanded to AI explainability~\cite{min_feedforward_2025,gmeiner2025intent,park2024coexplorer,park2024coexplorer_chi} and autonomous vehicles~\cite{sandhaus2018woz}.

Unlike feedback, feedforward supports discovery by revealing what is possible in advance~\cite{vermeulen_crossing_2013}. This makes it well-suited for surfacing hidden knowledge. However, poorly designed feedforward can mislead users~\cite{lafreniere2015these}, so systems should promote exploration rather than passive acceptance. While traditional feedforward communicates user interface interaction instructions (e.g., ``Slide to unlock''), we explore the methods to communicate constituent knowledge of common ground in virtual production instead of the interaction instructions, such as creative decision (e.g., ``The shelf should have silvery color'').

\subsubsection{In-situ Presentation of Information} 
A related strategy for surfacing hidden knowledge is the in-situ presentation of information, where relevant information is embedded directly in the user's workspace for just-in-time access~\cite{bressa2021s}. This helps users relate disparate pieces of information. In programming, inline visualizations help users understand running code~\cite{bianchi2024inline,hoffswell2018augmenting}; in text-rich contexts, embedded visualizations keep users focused on the text while providing information that helps interpret it~\cite{goffin2020interaction,sultanum2021text,wang2022phenopad}. Comments and annotations relevant to parts of 2D canvases~\cite{ellis2004collaborative,heer2009voyagers,kauer2024discursive} or video comments~\cite{chen2023visualizing,ji2025classcomet} can be placed on the relevant portions of the canvas or video, helping viewers interpret the material better and fostering collaboration. Commercial tools like Miro\footnote{\url{https://miro.com}}, Autodesk RV\footnote{\url{https://www.autodesk.com/products/flow-production-tracking/rv}}, and SyncSketch\footnote{\url{https://syncsketch.com}} embody these principles. We also explore annotations to encourage collaboration and knowledge acquisition, but we aim to explore communicating this knowledge in a 3D editing environment. 

In 3D and physical spaces, in-situ information can be projected to reveal context. Data associated with physical elements can be presented on physical spaces or objects, such as weather data or the star ratings of restaurants~\cite{willett2016embedded, lee2023design,caggianese2019situated}. Information relevant to conversations can be presented on the communicators' faces~\cite{pohl2024body,janaka2022paracentral} to obtain information without disrupting attention to the communicator. Tools relevant for thinking about objects, such as calculators~\cite{pohl2024integrated}, or annotations in physical spaces can be provided in-situ~\cite{numan2025cocreatar}. Users can use a virtual hand in a remote physical space to refer to a location, essentially acting as a real-time in-situ annotation~\cite{sodhi2013bethere,lee2014annoscape}, or a physical hand in a virtual space to provide real-time, in-situ annotations on their presentations~\cite{liao2022realitytalk,saquib2019interactive}. In virtual 3D canvases (similar to the domain of our work), such as architectural renderings, comments and annotations can be provided in-situ so that viewers can better associate the location in the 3D space with the annotations or comments~\cite{jung2002annotating,jung2001space}. In this setting, not just text, but multimedia comments can be included, such as audio or video, with tree-based representations~\cite{guerreiro2014beyond}. 

\added{Many of these systems treat comments or annotations as static, non-reactive marks attached to assets. For collaboration systems for film production, these comments are often also decoupled from the tools where changes are actually made. By contrast, our goal is to treat meeting-derived comments and decisions as a dynamic, in-situ feedforward layer inside the primary 3D editing tool. Rather than asking users to open a distinct review environment (e.g., annotations on the video clips, instead of 3D assets), infer the meaning, and be wary of not violating some constraints suggested, we proactively surface constraints and rationales directly as they manipulate objects, with each cue linked back to its originating meeting recording clip and decision entry. We aim to help users form common ground this~way. }

Particularly, we recognize that in-situ annotation can guide users, for instance, for movement~\cite{cheng2024viscourt,anderson2013youmove}. In this work, we combine feedforward with in-situ presentation in a 3D editor environment to guide users through the meeting context while equipping them with constituent knowledge for common ground formation. Through this, we aim to make collaborators more informed, aligned, and seamlessly connected with other collaborators. 

\section{Formative Study}
Although prior works have examined the workflows, advantages, operationalizations, and drawbacks of virtual production (Section \ref{sect:b-vpc}), few have investigated how virtual production staff collaborate, which leaves open questions (\textbf{RQ1}).
We conducted a formative study with virtual production professionals to address this.

\subsection{Method} 
We sent email invitations to a company internal research community consisting primarily of customers who had indicated their primary industry to be film or whose primary software expertise included 3D design or animation software. We recruited 23 participants\footnote{\textbf{Gender identity:} 21 M, 2 unknown; \textbf{Age:} 40.8$\pm$14.23; \textbf{Country of Residence:} 4 US, 4 India, 8 Canada, 1 Germany, 1 Honduras, 1 Spain, 1 South Korea, 1 Palestine, 1 UAE, 1 Zambia (Raw, typed inputs from participants.); \textbf{Years of experience in virtual production:} 6.2$\pm$4.18; \textbf{Number of past virtual production projects:} 28.2$\pm$44.98; \textbf{Roles:} 6 VP Artists, 4 VP Supervisors, 2 Software Developers/Engineers, 2 VP Coordinators, 2 Technicians, 1 VP Lead/Researcher, 1 VP Researcher, 1 Technical Director, 1 Animator, 1 Compositor, 1 VFX Supervisor, 1 Student}. Participants were invited to complete a Qualtrics survey \added{(\autoref{sect:formativeSQs})}, providing insights into their collaboration and onboarding experiences in virtual production. Some participants expressed interests in the follow-up interview \added{(See \autoref{sect:formativeQs} for the list of questions)}, which the aim was to help the researchers understand some of the unique practices mentioned in the survey, in more detail. We randomly selected 6 from the 18 that were interested (demographics: \autoref{tab:interview} (appendix)). 
Participants received gift cards for participation (survey: 15 minutes, US\$10; interview: 45 minutes, US\$60)\footnote{All studies were approved by our institutional ethics board.}.

\label{sect:thematicAnalysis}
We performed thematic analysis~\cite{braun2006using} on all survey and interview responses\footnote{Transcribed via Dovetail and verified by a researcher}. \added{Our analysis followed an inductive approach, focusing on participants' explicit descriptions of their experiences rather than applying a predefined coding frame. The lead author conducted the analysis. First, the lead author reviewed transcripts in full and noted initial impressions. Next, the author performed open, line-by-line coding in Dovetail (without using AI features), generating codes that captured meaningful units of text. The lead author then iteratively reviewed and refined the initial codes, merging, splitting, and discarding codes as needed. Related codes were grouped into candidate themes. Groups were color coded for easier access when extracting insights. Throughout this process, the lead author discussed the results of coding with co-authors on a daily basis, who provided feedback on code and theme definitions, but no second formal coding pass or inter-rater reliability was calculated, in line with reflexive thematic analysis. Finally, the lead author named and defined each theme and selected representative quotes to illustrate them in the findings.}
In this paper, \textbf{F$n$} denotes a unique formative survey participant. Some participants (\textbf{F2,5,10,15,17,22} - see \autoref{tab:interview}) also participated in formative study interviews, denoted as \textbf{FI$n$}.

\subsection{Collaboration in Virtual Production}

\begin{figure*}[!htb]
    \centering
    \includegraphics[width=\linewidth]{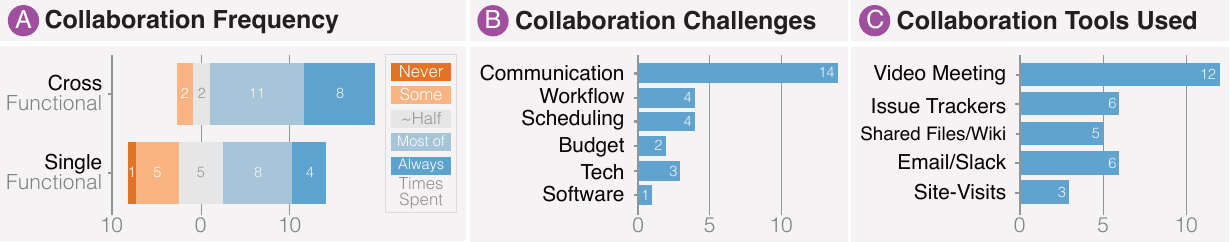}
    \caption{Formative study results on collaboration practices for virtual production. (A) shows frequency of collaboration for cross- vs. single-functional collaboration. (B) shows the challenges in current collaboration practices (multiple responses possible). (C) shows the tools that professionals currently use for collaboration (multiple responses possible).}
    \Description{The figure has 3 bar charts. A shows a stacked bar chart of cross-functional and single-functional collaboration frequency. B shows a bar chart of collaboration challenges. C shows a bar chart of tools used in collaboration.}
    \label{fig:formative_colab}
\end{figure*}

Collaboration in virtual production is extensive and cross-functional. The majority of participants work frequently with colleagues from other roles (19/23; \autoref{fig:formative_colab}A-top), notably more often than with those in their same role (12/23; \autoref{fig:formative_colab}A-bottom). However, this collaboration practice brings challenges, as \textbf{F15} illustrates: \textit{``Issues can arise weekly or even daily during fast-paced phases like scene assembly or on-set shooting, especially when multiple departments are involved.''} Most often, these issues relate to communication (\autoref{fig:formative_colab}B).

\subsubsection{The Root Cause: Lack of Common Ground}
Our analysis revealed that communication issues primarily stem from insufficient formation of common ground. The short history combined with complexity of collaboration in virtual production results in a lack of standards, as \textbf{F15} summarized: \textit{``Because VP is still evolving, not everyone shares the same level of experience, which can lead to misunderstandings and uneven expectations.''}

Specifically, 12 participants experienced issues indicating the factors, attributable to the \textbf{insufficiency in a formed common ground} as the main cause for communication problems. These issues encompassed inconsistencies in nomenclature, versioning style, intended workflow, tacit technical/artistic limitations, styles, and managing expectations. This lack of shared understanding manifests in several~ways:

\textbf{Unclear Intent Behind Actions.} \textbf{FI1}, \textbf{FI2}, and \textbf{FI5} experienced difficulties understanding the \textbf{intention behind collaborators' actions}. When intent is unclear, \textbf{FI2} improvises: \textit{``We know the bare minimum that we need to know [about the film/script]. But we are also given quite a lot of freedom to improvise for whatever we need to film or perform.''}

These communication gaps result in production frictions with real consequences, as \textbf{FI1} explains: \textit{``An artist will be working on [a 3D scene] and then they'll inadvertently grab something and move it without realizing it. It's happened to me many times, and then all of a sudden, [when we load] everything is out of calibration. Nothing is lining up and we have to troubleshoot.''} Such misalignments are particularly problematic when dealing with ``Hero Assets'', which are assets important for the narrative or immersion in the scene (\textbf{FI4}). Poor coordination can even pose physical hazards, as \textbf{FI2} explains: \textit{``The camera unfortunately came too close to [the actors] and he was spinning this whole [camera setup]. He knocked off the lens.''}

\textbf{Information Propagation Failures.} For 4 participants, \textbf{information propagation} was the main communication issue. They do not obtain information about the most recent decisions or changes. For 3 participants, \textbf{unclear explanations of responsibilities and expectations} were the source of problems. These failures result in added production costs, as \textbf{FI1} illustrates: \textit{``This last-minute change is going to affect everything else that we planned for and we built for. And it would have been great if you had told me that this was an option so that we could have been prepared to make those adjustments. So sometimes, the departments operate in a vacuum.''}

\textbf{Inadequate Knowledge Resources.} Despite the availability of shared materials (\autoref{fig:formative_colab}C), some professionals experience difficulties in establishing the common ground (e.g., \autoref{fig:formative_onboarding}C). \textbf{FI6} described using internal Wiki pages and style bibles containing guidelines for each scene and project. However, \textbf{FI6} observed that collaborators often do not engage with these resources: \textit{``I sometimes doubt that people actually look at all the information wiki pages because a lot of times we do get questions or we do get tickets for it or for another department. [Collaborators not considering the style guidelines] is more with junior artists and that comes up in dailies or during a review time or something. [Some people are like] people who don't read the instruction manuals and just hope that they can wing it.''}

From these findings, we derive a design goal:

\aptLtoX{\begin{framed}
\textbf{Design Goal 1 (DG1):} 
Gathering and making
constituent knowledge accessible which is needed for common ground such as intention behind actions, constraints, responsibilities, nomenclature, changes, and version history.
\end{framed}}{\vspace{2mm}
\noindent\fbox{%
   \parbox{0.47\textwidth}{
\textbf{Design Goal 1 (DG1):} 
Gathering and making
constituent knowledge accessible which is needed for common ground such as intention behind actions, constraints, responsibilities, nomenclature, changes, and version history.}}}

\subsubsection{Meeting-Driven Resolution with Limitations}
When collaboration issues arise, participants typically rely on (mostly remote) meetings (12/23) to resolve them (\autoref{fig:formative_colab}C). Issue trackers like Jira or ShotGrid, as well as shared documents like Notion, Google Docs, or Confluence are also commonly used. Convenience drives this preference, as \textbf{F10} notes: \textit{``We use Zoom meeting since the team is all over the world''} and \textbf{FI1} states: \textit{``
It's just easier to get everybody in one room [in Zoom].''} \textbf{FI3}, \textbf{FI5}, and \textbf{FI6} used AI solutions to generate summaries of meeting recordings. \textbf{FI5} explained the method: \textit{``We'll take the transcript and use ChatGPT to generate different versions of summaries. Strategy there is pretty basic.''}

Some participants take proactive approaches to surface potential issues. \textbf{FI1} conducts daily reviews to catch issues early: \textit{``We always try to have a daily, just to make sure if there were changes, how they're looking. I always want to do an A/B comparison. I always want to compare what it was and what we did, what changes we made. So, if there's things that got changed, I'd rather want to know about it before we load it into the volume stage and start rolling cameras.''}


\aptLtoX{\begin{framed}
\noindent\textbf{Design Goal 2 (DG2):} Proactively deliver relevant knowledge (instead of just acting as a reference place for the knowledge) from meeting histories at the moment of need during professionals' work.\end{framed}}{\vspace{2mm}
\noindent\fbox{%
   \parbox{0.47\textwidth}{
\textbf{Design Goal 2 (DG2):} Proactively deliver relevant knowledge (instead of just acting as a reference place for the knowledge) from meeting histories at the moment of need during professionals' work.}}\vspace{2mm}}

\subsection{Onboarding Collaborators to a 3D Scene} 
\label{sect:formative-onboarding}

\begin{figure*}[!h]
    \centering
    \includegraphics[width=\linewidth]{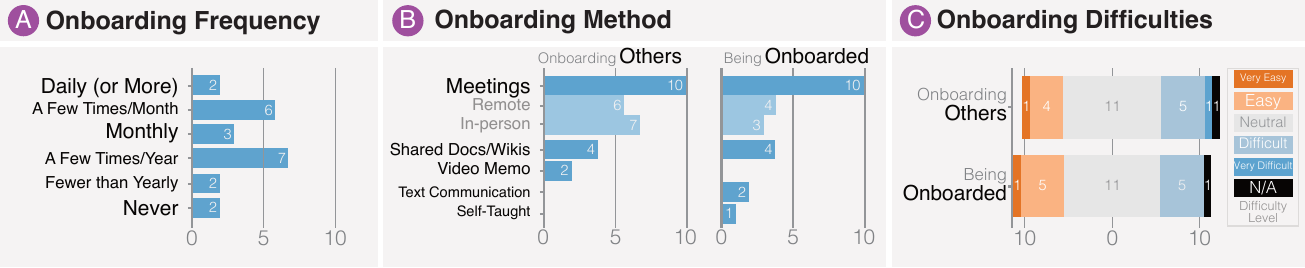}
    \caption{Formative study results on onboarding practices in virtual production. (A) shows the frequency of the onboarding activities, (B) shows the methods that collaborators use (multiple responses possible - each is out of 15/23 who reported their onboarding experiences.), and (C) shows the self-reported difficulty levels of onboarding, from the perspectives of onboarding others and being onboarded to a scene.}
    \Description{The figure has 3 bar charts. A is a bar chart showing the frequency of onboarding in virtual production. A few times per year has the highest frequency. B is a bar chart showing the method of onboarding used, with meetings being the most frequent. C is a stacked bar chart showing the difficulty level of onboarding others and being onboarded.}
    \label{fig:formative_onboarding}
\end{figure*}

\noindent \textbf{Onboarding as a Significant Collaboration Challenge.} We define ``onboarding'' as not merely an activity of joining a whole new film project, but rather a more routine activity of bringing a new collaborator up to speed on a scene in a project, particularly regarding the 3D scene used for virtual production. Due to the project-driven employment nature of film production, we predicted that such onboarding would be frequent. Our findings support this (\autoref{fig:formative_onboarding}A). Since staff being onboarded start without common ground, effective onboarding is critical for project success.

\noindent \textbf{Inconsistent Onboarding Experiences Reveal System Gaps.} \autoref{fig:formative_onboarding}B shows that remote video meetings are commonly used for onboarding, regardless of whether participants are onboarding others (6/15) or being onboarded themselves (4/15). However, experiences vary dramatically. While \textbf{F6} prepared personalized materials (e.g., video memos) which may help the worker being onboarded, they themselves reported that onboarding remains very difficult. Some participants perceived both being onboarded and onboarding others as challenging, with bi-polar responses overall (\autoref{fig:formative_onboarding}C).

The quality of onboarding depends on individual organizational skills and project management, as \textbf{FI1} describes: \textit{``We'll rewatch the lensing sessions together. We'll look at certain things and then we'll make notes. Typically, I do it because I like taking notes and have to make sure that my stuff is on point.''} However, even experienced professionals struggle with poorly organized projects. \textbf{FI1} explains: \textit{``I've worked on projects that have been so, so disorganized. It's frustrating. You're wasting so much time, you're wasting so much money, and nothing ever gets done. [...] On disorganized projects, we have to start the QA process weeks ahead of time.''}

\textbf{FI3} highlights how inadequate onboarding affects production quality: \textit{``In medium and small productions, untalented professionals would make a huge setback for the whole set. Because the whole time you are trying to fix things and figure things out in a production time. It has to be a preparation time before.''}

Most concerning, onboarding is often minimal or nonexistent. \textbf{F23} noted: \textit{``Honestly, onboarding is a luxury—usually a project is delivered and it's up to the team to ingest and learn the project well enough to run it during production.''} \textbf{FI5} describes a common approach: \textit{``[For onboarding,] I usually just give them the file. Dig around, and figure it out. Let me know if you have questions.''}

\aptLtoX{\begin{framed}
\noindent\textbf{Design Goal 3 (DG3):} Automate common ground establishment to ensure consistent, effective onboarding experiences across all projects, reducing dependency on individual organizational skills and manual knowledge transfer processes.
\end{framed}}{\vspace{2mm}
\noindent\fbox{%
   \parbox{0.47\textwidth}{
\textbf{Design Goal 3 (DG3):} Automate common ground establishment to ensure consistent, effective onboarding experiences across all projects, reducing dependency on individual organizational skills and manual knowledge transfer processes.
}
}
\vspace{2mm}}
\section{\emph{GroundLink}}

We designed \emph{GroundLink}, a Unity add-on,\footnote{We implement \emph{GroundLink} as a Unity add-on. We chose Unity because it is one of the game engines used for virtual production and is comparatively easy to learn~\cite{zwerman_ves_2023}.} to operationalize our design goals (\textbf{DG1–DG3}) and to study their impact on virtual production professionals' common ground formation (\textbf{RQ2}) and confidence/ease when editing 3D scenes~(\textbf{RQ3}).

\begin{figure*}[!htb]
    \centering
    \includegraphics[width=\textwidth]{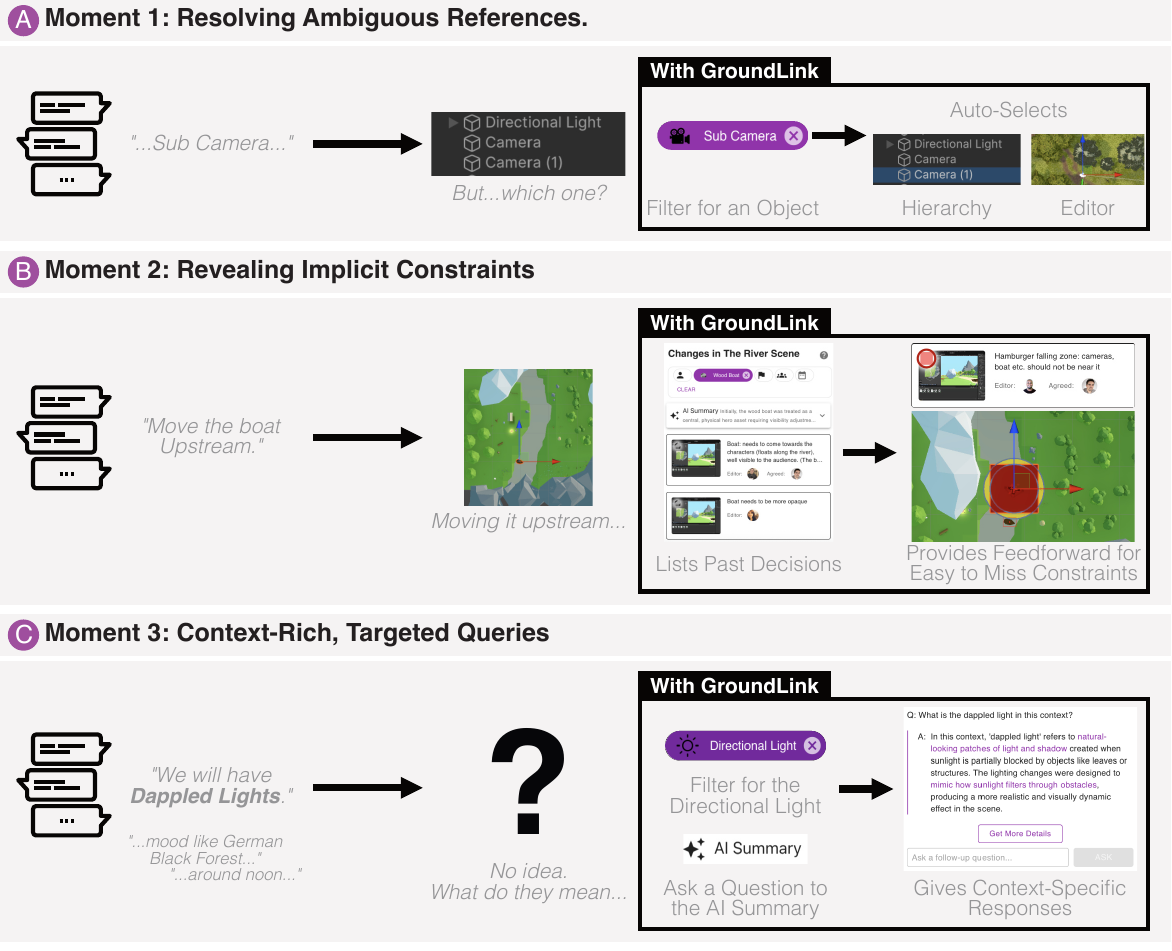}
    \caption{To illistrate how \emph{GroundLink} supports common ground formation, consider Bob, a VFX artist newly assigned to a river scene. Traditionally, Alice (the production supervisor) would onboard him through meetings and explanations of past decisions. With \emph{GroundLink}, Bob instead uncovers the scene's collaborative history while working. \\ \textbf{(A) Moment 1: Resolving Ambiguous References.} Bob is tasked with moving the sub camera. With \emph{GroundLink}, clicking the object in the dashboard highlights the specific camera in the 3D editor, eliminating the need for second guessing. \\ \textbf{(B) Moment 2: Revealing Implicit Constraints.} Bob's task is to move the boat upstream, but he does not know a no-placement zone was established two meetings ago. With \emph{GroundLink}, past decisions and a feedforward cue appear as subtle red zones in the editor, expanding and darkening as the boat nears. This helps Bob avoid placing the boat in the restricted area. \\ \textbf{(C) Moment 3: Context-Rich, Targeted Queries.} Bob is tasked with adjusting the light to create a dappled light effect, a term he does not know or understand in this scene. With \emph{GroundLink}, he can ask the AI summary a targeted question about the scene and object of interest, avoiding unnecessary second guessing.}

    \Description{A shows the moment of resolving ambiguous references, where the speech bubble reads ``…sub camera...''. Then, there is a box showing multiple objects with the text but which one?’’. With GroundLink, Bob can filter for an object, which auto-selects the hierarchy and editor. B shows the moment of revealing implicit constraints. The speech bubble reads ``move the boat upstream’’, with Bob trying to move it upstream. With GroundLink, it lists past decisions and provides feedforward for easy-to-miss constraints in the editor, avoiding the restricted zones when placing the boat. C shows the moment where context-rich, targeted queries become useful. The speech bubble reads ``we will have dappled lights’’, which Bob does not understand. With GroundLink, Bob can filter for that object and ask a question to the AI summary, which gives context-specific responses.}
    \label{fig:scenario}
    \vspace{-1.5em}
\end{figure*}

\emph{GroundLink} detects user interactions in the Unity editor and identifies when edits diverge from prior decisions captured in meeting recordings (\autoref{fig:teaser}). It then surfaces in-scene feedforward cues in the editor to reveal potential consequences of the current action, and simultaneously highlights the linked decision/comment on the dashboard timeline while seeking the meeting video to the exact segment. The dashboard, editor, and feedforward cues are bidirectionally synchronized (example scenarios: \autoref{fig:scenario}).


\vspace{-1em}

\subsection{Meeting Knowledge Dashboard for Capturing and Reviewing Decisions and Comments (DG1)}

\begin{figure*}[!htb]
    \centering
    \includegraphics[width=\linewidth]{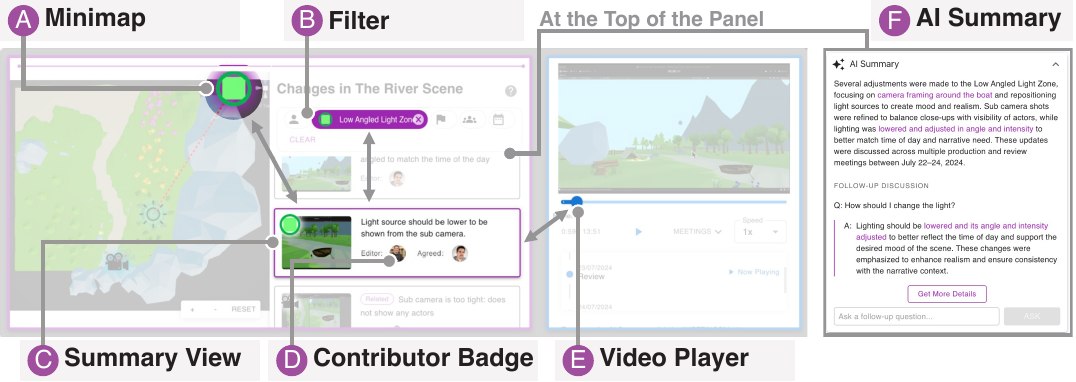}
    \caption{\emph{GroundLink} summarizes verbal information in a sequential representation on a dashboard. (A) The minimap displays all element changes on a zoomable map. (B) Users can filter decision/comment entries by collaborator, object, meeting phase, meeting, and date range. (C) The summary view lists all changes/comments. (D) For each entry, contributor badges and agreements are attached. (E) The video player with timeline (synchronized with other panels) and meeting list allows users to watch video while reviewing decisions/comments. (F) The AI summary and chat enables quick glances and Q\&A.}
    \Description{The figure shows the meeting knowledge dashboard. It has a minimap on the left, a summary view in the middle, and a video player on the right.}
    \label{fig:dashboard}
\end{figure*}

\emph{GroundLink} provides a temporal dashboard that consolidates meeting-derived, verbal knowledge, and meeting recordings in one place (\autoref{fig:teaser}E). A zoomable minimap displays changes across meetings as a cohesive spatial overview; a filterable, chronologically ordered list captures decisions and comments with identities of editors and agreeing collaborators; and an integrated video player anchors each entry to its source segment (\autoref{fig:dashboard}). In doing so, the dashboard directly operationalizes \textbf{DG1} by gathering constituent knowledge needed by VP professionals.

Unlike prior production dashboards\footnote{e.g., Autodesk Flow (https://www.autodesk.com/company/autodesk-platform/me) or Perforce (https://www.perforce.com/products/helix-core)}, entries in \emph{GroundLink} synchronize with the rest of the application. Each entry has a mapping to an object, part of a 3D scene, and a video meeting snippet, binding every insight to concrete objects, moments, and collaborators. Filtering is likewise object, scene, and meeting-based, and the AI summary operates over exactly the currently scoped subset, yielding focused, on-point synthesis.

The \underline{minimap} (\autoref{fig:dashboard}A) visualizes an object's location history as a dashed path and emphasizes selected elements. Selecting an object on the minimap filters the summary view to that object and highlights the corresponding entries (\autoref{fig:dashboard}B–C). All elements in the minimap and list are filterable by object, collaborator, production phase, meeting instance, and date range; when filters change, the minimap emphasizes relevant traces by increasing opacity and putting purple glow behind relevant objects while de-emphasizing others.

The \underline{summary view} (\autoref{fig:dashboard}C) presents significant production decisions and comments in chronological order, with a thumbnail of the linked video snippet and a badge for the related scene element. Each entry shows a concise summary plus the editor and agreeing collaborators (\autoref{fig:dashboard}D). Hovering over a badge reveals role, name, and, for agreements, any qualifying remarks. These entries can be filtered (\autoref{fig:dashboard}B).

On the right, the \underline{video player} (\autoref{fig:dashboard}E) supports rapid verification: clicking any summary entry seeks to the exact moment in the meeting. There is a one-to-one correspondence between list entries and meeting segments.

A \underline{collapsible AI summary} (\autoref{fig:dashboard}F) sits above the list and respects the current filters (\autoref{fig:teaser}F). Users can obtain scoped syntheses (e.g., only the ``shelf'' object or a collaborator ``Alice''), ask follow-up questions, and jump via highlighted references to the corresponding list entries. This serves both as citations and as navigational~shortcuts.

\added{The dashboard is designed to go beyond making individual decisions more trackable, by revealing higher-level patterns about how the team works. The minimap and filters reveal how constraints accumulate over meetings, in what area in the scene, the changes happen, which movie departments tend to make particular changes/suggestions, and how often decisions are revisited. Contributor badges expose who usually edits or agrees to a given element.}

We draw design rationale from Dual Coding Theory~\cite{paivio1991dual}. Dual Coding Theory posits partially independent verbal and non-verbal systems for interpreting two types of information. The dashboard preserves the optimal sequential representation for verbal artifacts (time-coded transcripts, decisions, comments), while other views maintain in-situ, visual representations of scene state.

\subsection{Constraint-Aware Feedforward for Informing Creative Decisions and Comments (DG2)}

\begin{figure*}[htb]
    \centering
    \includegraphics[width=\linewidth]{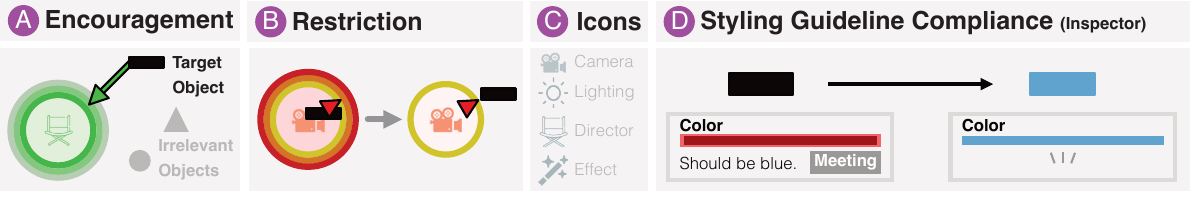}
    \caption{\emph{GroundLink} surfaces in-situ feedforward derived from meeting recordings: (A) encouragement zone, (B) restriction zone, (C) department icons, (D) inspector feedforward for styling compliance.}
    \Description{The figure shows constraint-aware feedforward mechanisms. A shows an encouragement zone, which is represented as a green circle with a director icon in the middle. There is an arrow coming from the target object toward the zone. B shows a restriction zone. If the target object is inside the zone, there are red, orange, and yellow rings around the zone, represented as a circle. When the object leaves the zone, the colored rings disappear. C shows the icons used for the zones, which are camera, lighting, director, and effect icons. D shows styling guideline compliance feedforward, which appears in the inspector. If the target object is black but should be blue, the color field blinks red, with the description text below reading ``should be blue.'' A meeting button appears on the right-hand side of that. If the color is changed and becomes blue, that feedforward disappears.} 
    \label{fig:feedforward}
\end{figure*}
Operationalizing \textbf{DG2}, during direct manipulation in the Unity editor, \emph{GroundLink} overlays constraint zones (e.g., areas to avoid, preferred placements) as in-situ, visual feedforward so that relevant prior decisions appear during work rather than after-the-fact (\autoref{fig:teaser}A). In the inspector, when a property diverges from a previously agreed guideline (e.g., a color choice), the corresponding control draws subtle attention (\autoref{fig:teaser}D), presents a brief rationale, and offers a one-click jump to the linked decision on the dashboard.

Unlike existing tools that deliver post-hoc or cross-modal feedback (e.g., textual warnings about a 3D placement)~\cite{long2025feedquac,mandorli2024improving}, \emph{GroundLink} provides in-situ, same-modality feedforward: cues are rendered directly in the 3D editor at the locus of manipulation and update continuously as the edit unfolds. Classic feedforward communicates a definitive action (e.g., ``slide to unlock'')~\cite{vermeulen_crossing_2013}, whereas decisions encoded in meeting artifacts are inherently uncertain and context-dependent. If the system could infer the exact edit, it would simply apply the change and request a simple ``accept''; in practice, creative judgment remains essential. \emph{GroundLink} therefore treats ambiguity as a design resource~\cite{gaver2003ambiguity}: it modulates opacity, animation, and band thickness of circles to convey proximity, and links each cue to its provenance in the meeting record for immediate verification on the dashboard. This predictive, same-modality guidance with explicit uncertainty and provenance aims to reduce context switching while avoiding false feedforward~\cite{lafreniere2015these}.

Positional constraints are communicated with circular zones that appear only in the editor view\footnote{I.e., not on the camera view} (\autoref{fig:teaser}A). Encouragement zones (\autoref{fig:feedforward}A) use a green ring and a subtle arrow from the relevant object toward the target area; as the object approaches, the ring's opacity and animation intensity diminishes and banding simplifies, becoming transparent when the object is properly within the zone. Restriction zones (\autoref{fig:feedforward} 
B) convey the opposite gradient: salience increases as the object nears the restricted area and fades with distance. Each zone includes an icon at its center indicating the movie department that proposed the constraint (\autoref{fig:feedforward} 
C), supporting transactive memory about the source of guidance. Interacting with a zone reveals the corresponding entry on the dashboard's summary list and seeks the meeting video to the linked segment. The summary view also shows history of decisions and comments made for~the~zone.

Styling constraints are surfaced in the inspector (\autoref{fig:teaser} 
D). When a user selects a value that conflicts with a prior decision (e.g., having a black ``shelf'' when the agreed color is blue), the relevant control (e.g., color picker or opacity slider) gently blinks, accompanied by a short explanation (``Should be blue'') and a meeting button that navigates to the originating discussion on the dashboard (\autoref{fig:feedforward} 
D). In all cases, \emph{GroundLink} provides uncertainty-aware feedforward tied to verifiable sources, surfacing knowledge at the moment of need while preserving creative agency.
\vspace{-1mm}

\subsection{Cross-Modal Synchronization for Linking Meetings and Editor Scene Context (DG3)}
\emph{GroundLink} keeps the dashboard, the Unity editor's hierarchy, the scene view, and the video player synchronized (\autoref{fig:teaser}B). Unlike conventional production dashboards that isolate comments, meeting recordings, and the 3D editor, \emph{GroundLink} creates fine-grained, cross-modal links between timestamped meeting snippets, change/comment summaries, and concrete 3D objects. These links transform ambiguous references (e.g., ``sub camera,'' ``this zone here'') into unambiguous selections in the scene. 

Users no longer have to guess which scene element corresponds to which meeting artifact, or vice versa. This synchronization happens proactively, operationalizing \textbf{DG3} by automating common ground establishment during work. This means that if a user clicks or interacts with any part of the dashboard, the rest of the dashboard (minimap, summary list, filter, and the video player) and the editor (object selection in the hierarchy and zooming into the object in the editor) are synchronized, and vice versa.




Dual Coding Theory posits partially independent verbal and non-verbal systems, with people making referential connections between two systems~\cite{paivio1991dual}. Production workflows often already separate the channels e.g., verbal artifacts in notes/transcripts and non-verbal artifacts in the scene, but leave the referential links implicit. This forces users to perform mental indexing across channels, especially when terminology is inconsistent or spatial configurations have changed. \emph{GroundLink} preserves each channel's natural representation (sequential, time-coded verbal material; spatial, in-situ visual state) while externalizing the referential links between them. By turning phrases like ``the hero light'' into concrete, navigable selections and timepoints, the system reduces the burden of cross-modal mapping and promotes faster, more reliable interpretation without collapsing one channel into the other.

To further support \textbf{DG3}, the same mapping works in tandem with the feedforward that surface relevant prior discussion at the moment of need. For example, if a property mentioned in meetings is out of compliance, the corresponding inspector field draws attention and provides a one-click jump to the associated insight in the dashboard. This resembles the concept of ``information scent'' that encourages follow-up rather than interruption~\cite{chi2001using}.

\subsection{Implementation}
\added{\emph{GroundLink} operates in two phases. The main phase consists of realtime interaction, which is the focus of this paper and is used for the user study. For this to work, another phase is required, which is a onetime preprocessing phase. While essential for the system, this preprocessing phase is not the main focus of this paper and some part is implemented using a Wizard-of-Oz method. We explain each phase in detail.}

\subsubsection{\added{Realtime Interaction}}

\begin{figure*}[!htb]
    \centering
    \includegraphics[width=\linewidth]{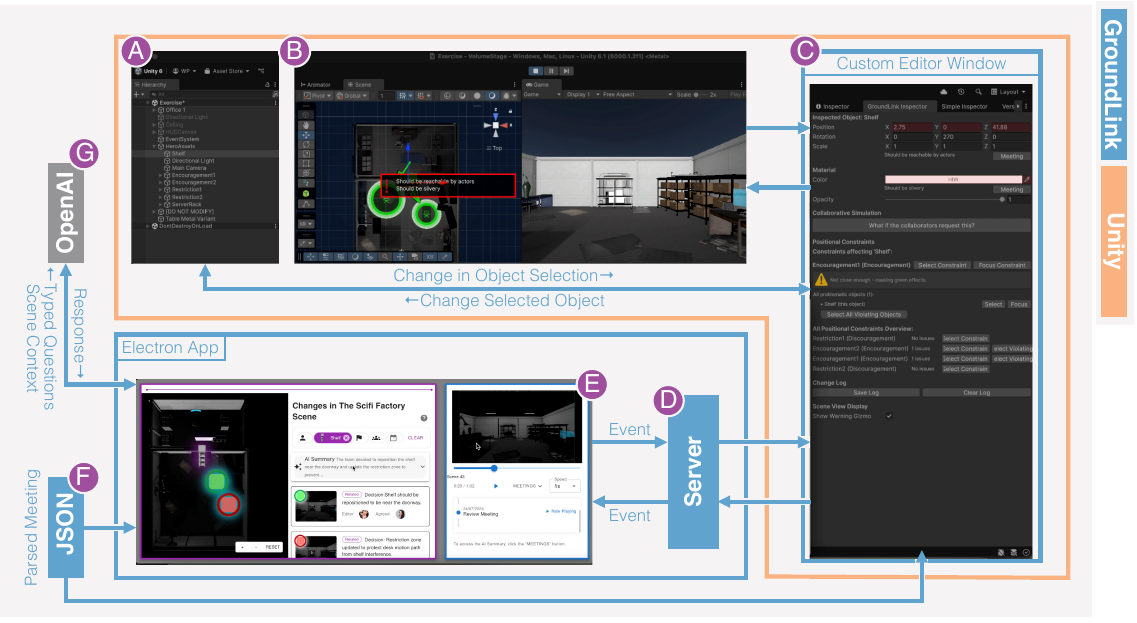}
    \caption[width=\linewidth]{In the realtime interaction phase, a (junior) worker works in Unity (Figure \ref{fig:teaser}). In Unity, there is a hierarchy view (A), scene/game view (B), and \emph{GroundLink} is integrated on the (C) as a Unity custom Editor window. This window collects interactions from other Unity views (e.g., selecting a game object, changing the position/property of an object) and can also modify them (e.g., select another game object, put feedforward effects). To do this, the window communicates with a server (D) implemented in an Electron app via WebSocket. The same server also communicates with a React-based application (E), which interprets Unity changes and fetches updates to the Unity custom Editor window. In doing so, it uses the JSON (F) generated in the preprocessing phase. The React-based application uses OpenAI's API (G) to provide the real-time AI features. Note: The tools/processes we developed are marked with blue color, whereas the existing services/tools are colored in orange. 

    }
     \label{fig:system}
    \Description{JSON is fed into GroundLink, which is an Electron app. The same app also has a server, and it communicates with GroundLink in Unity.}
\end{figure*}

\added{Figure 7 illustrates how the two components, which comprise the add-on are configured:
(1) an Electron}\footnote{v36.5.0; https://www.electronjs.org}\added{ application built with React}\footnote{v19.1.0; https://react.dev}\added{ and Material UI}\footnote{v.7.1.1; https://mui.com}\added{ that hosts the temporal dashboard and video player, and (2) Unity}\footnote{v6.1 (6000.1.2f1)}\added{ custom Editor windows}\footnote{https://docs.unity3d.com/6000.2/Documentation/Manual/UIE-HowTo-CreateEditorWindow.html}\added{ that render in-editor/inspector feedforward, and log interactions. The components communicate via WebSockets, with a Node.js}\footnote{v22.16.0; https://nodejs.org/en}\added{ server embedded in the Electron app. It publishes selection, filter, and playback events and editor state changes. Concretely, the Unity custom Editor window runs in the editor update loop and monitors the current editor state (e.g., selected game objects and inspector properties). On each update, it compares this state to a cached snapshot and, when it detects a change, emits an event describing the interaction. These events are streamed over the WebSocket connection to the server, which forwards them to the dashboard application and, when appropriate, requests summaries and explanations from the AI service (Real-time AI features through OpenAI's GPT-5-chat API}\footnote{https://platform.openai.com/docs/models/gpt-5-chat-latest}\added{). In parallel, the Unity Editor window checks the updated state against a JSON file containing the constraint information and triggers feedforward effects in Unity by creating or removing game objects that visualize these effects. When the user clicks elements on the dashboard, it emits events back through the server to the Unity Editor window, which updates the Unity state (e.g., changing the selected object or zoom level), keeping the scene view and dashboard in sync. We designed }\emph{GroundLink}\added{ with a two-component architecture to help users continue working with the content-creation tools they are already familiar with, while also providing advanced features such as integrating video APIs, AI APIs, data import, and animation effects, not supported by Unity or difficult to implement~with~Unity.}

\subsubsection{\added{Onetime preprocessing}}

\begin{figure}
    
    \centering
    \includegraphics[width=\linewidth]{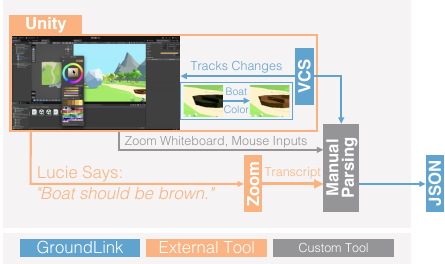}
    \captionsetup{type=figure}
    \caption{\added{During meetings, the Version Control System (VCS), a component of \emph{GroundLink} implemented as a Unity custom Editor window, tracks changes in the scene, such as changes to transforms (position, scale, rotation) and properties (color, opacity, field of view, etc.). Zoom Whiteboard/mouse inputs are also collected (manually decoded), and speech from the Zoom meeting is transcribed through Zoom and collected. All collected data is manually matched by the lead researcher using an Authoring Tool, a custom-made web app that presents the changes and helps match changes and mentions. This process generates JSON files. The JSON generated is then used in the next phase. Note: Grey colored part is also developed by us, but is not the focus of this paper.}}
    \label{fig:systemSecond}
    \Description{Unity has a VCS, and it tracks changes. Information from Unity and the VCS flows into the Authoring Tool, which then flows into JSON.}
    \vspace{-2em}
\end{figure}

For the user study, we used pre-indexed meeting recordings with timestamped transcripts, and scoped our exploration to knowledge consumption rather than parsing them realtime. We therefore manually annotated transcript-to-scene mappings to isolate interaction design effects rather than confound results with automatic NLP accuracy. \added{Figure \ref{fig:systemSecond} illustrates this onetime, preprocessing. This is done through a parsing process we developed, which the researchers match data from different sources}\footnote{Transcripts of the recordings, Zoom annotations including whiteboard and mouse movements, changes in the Unity Editor}\added{ by manually clicking the instances of change or snippets of transcripts}\footnote{We developed a basic web application tool for this.}. 
\added{We discuss how the parsing process}\footnote{We understand that the parsing process, which is part of a broader authoring process, is an important part of the system, despite not being the focus of this paper.}\added{ could be implemented in future work, later in the paper (\autoref{sect:doubleEdgedSword}). This process of manually matching information from different sources can be characterized as a Wizard-of-Oz (WOz) method. This strategy allows researchers to build systems when current technology falls short of meeting usability requirements, so that evaluation of the rest of the system}\footnote{Which is the scope of this paper.}\added{ can be conducted~\cite{kelley1983empirical,maulsby1993prototyping}. WOz is a popular, validated method in the human-AI interaction community~\cite{leong2024dittos,gmeiner2025exploring,hu2023wizundry,dow2010eliza,teng2024human}.} 

\section{Exploratory Comparative User Study and Expert Evaluation}

We conducted two complementary studies to evaluate \emph{GroundLink}'s effectiveness in supporting virtual production workflows, with emphasis on newcomer experience but beyond as well. First, we performed a comparative user study (N=12) to assess how meeting-derived knowledge affects users' ability to form common ground with remote teams and complete VP tasks. Second, we conducted an expert evaluation (N=5) with VP professionals to gather domain-specific insights. Together, these studies address our research questions about common ground formation (\textbf{RQ2}) and task confidence (\textbf{RQ3}). Both studies received review and approval from our institution's internal ethics review process. Participants received a US\$100 gift card for their participation, for each study. 

\subsection{Exploratory Comparative User Study}
The goal of the comparative user study was to evaluate how meeting-derived knowledge presented through \emph{GroundLink} can help or hinder users' virtual production workflows, and whether it helps users perceive that they form common ground with existing remote teams. 

\subsubsection{Participants}
We recruited 12 participants\footnote{\textbf{Age:} 33$\pm$10.6 (mean; one did not report); \textbf{Gender:} 7 M, 4 W, 1 Unknown; \textbf{Occupation:} 1 post-doctoral researcher, 2 research scientists, 1 software developer, 1 associate research \& design engineer, 1 software development manager, 1 senior software developer, 2 software engineers, 3 user experience designers.} (completely new) through an internal mailing list. Unlike the formative study, we recruited participants with experience using 3D manipulation tools (Unity, Unreal Engine, Maya, or similar)\footnote{5 participants had used the tools for architecture, 7 for media and entertainment, and 6 for manufacturing. Participants had an average of 5.1$\pm$3.92 years of experience with the tools.}. We broadened eligibility to reach participants with more diverse backgrounds as virtual production is a growing area with people from diverse areas of expertise joining the industry~\cite{kadner_noah_virtual_2021,zwerman_ves_2023}. Therefore, this study sought to evaluate the capabilities of the system in isolation, while offloading VP-specific validation to the expert evaluation.

\subsubsection{Materials and Methods}
\begin{figure*}[!htb]
    \centering
    \includegraphics[width=\linewidth]{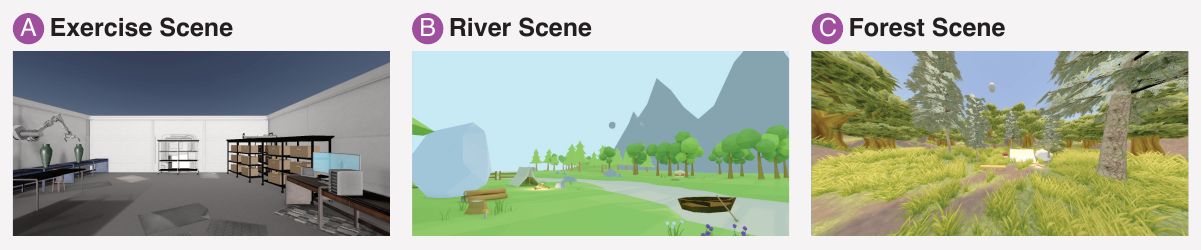}
    \caption{Three pre-assembled scenes used in the study: (A) exercise scene (a sci-fi office) that was used for demonstrating functionality to participants, (B) river scene in which the actors plan to use a boat to escape from food falling from the sky, only to realize the boat is an illusion, and (C) forest scene where actors eat fruit and talk around a campsite, followed by a monster appearing from the forest.}
    \Description{The figure shows three assembled 3D scenes. A shows a sci-fi-styled office, B shows a river scene with a campsite and a wooden boat on the river, and C shows a forest scene with a campsite in a forest.}
    \label{fig:scenes}
\end{figure*}

To implement the design concepts in a prototype, we created three virtual production scenes (\autoref{fig:scenes}). 
These scenes were co-created by two professionals with filmmaking experience working with two of the authors; these professionals did not take part in any other survey, interview, or study. We involved filmmakers to provide ecological validity in the recordings. This co-creation process resulted in three Zoom\footnote{\url{https://www.zoom.com}} video meeting recordings per scene\footnote{River and forest scenes only. For the exercise scene, we scripted the meeting and simulated the audio using ElevenLabs Text-to-Speech (\url{https://start.elevenlabs.io}). \added{We scripted the exercise scene meeting to provide a clear, controlled illustration of the system's functionality within a short period of time. These scripted meetings helped us efficiently cover all the features.}}, each approximately 15 minutes long. The recordings were semi-scripted, enabling comparable scenes with similar difficulty and level of detail for the edit process. For the river and forest scenes, the recordings established prohibited zones\footnote{Actor zone where cameras should not go; hamburger-falling zone for the river scene; and a monster-appearance zone for the forest scene where no elements should enter.}, styling guidelines (e.g., opacity and color), and target zones for camera or other elements (e.g., tree, mushroom). The recordings used mouse cursor circling\footnote{Moving the mouse cursor in a circular motion around the point of interest to emphasize.}, Zoom whiteboard annotations, and placement of placeholder cubes in the scene to indicate approximate size and position of each zone, alongside verbal descriptions of styling guidelines. Researchers ensured that each of the aforementioned methods was present in both sets of recordings.

After the recordings, one of the researchers manually parsed the contents, adding the mentioned zones in Unity and encoding styling guidelines in a JSON file. These scene ``constraints'' had a 1:1 mapping with discrete comment/change items from the recordings; each item was identified and encoded in a JSON file.

We installed \emph{GroundLink} with the scenes on an Intel Mac running macOS\footnote{macOS 15.6; CPU: 2.4 GHz 8-Core Intel Core i9; RAM: 32 GB; GPU: AMD Radeon Pro 5500M 8 GB}. Participants took part in an in-person study using this computer. The study lasted approximately 90 minutes.

\subsubsection{Task}
\begin{figure*}[!htb]
    \centering
    \includegraphics[width=\linewidth]{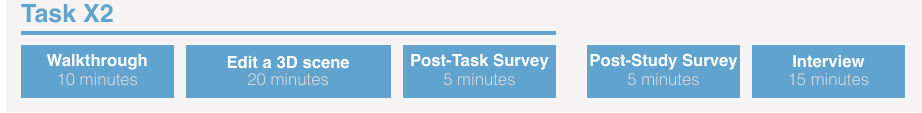}
    \caption{Task sequence for the comparative user study. It has within-subjects design with counterbalanced conditions.}
    \Description{The figure shows the task sequence. There is a task that repeats twice. In the task, there is a walkthrough (10 minutes), an edit a 3D scene (20 minutes), and a post-task survey (5 minutes). After the task, there is a post-study survey (5 minutes) and an interview (15 minutes).}
    \label{fig:task}
\end{figure*}

We selected Microsoft Copilot\footnote{\url{https://copilot.microsoft.com}} on ClipChamp\footnote{\url{https://clipchamp.com/en/}} as the baseline (with GPT-5 mode enabled), as it represents current best practices in AI-assisted meeting review. The baseline supported raw video meeting recordings (with standard video playback controls including playback speed), AI summary and chat about the video (with links to specific video segments from the summary), transcript, and search features. As reported in our formative study (\autoref{sect:formative-onboarding}), scene onboarding in VP commonly occurs through meetings, shared documents, and video memos. While the typical baseline would be simple video players with transcripts/notes, we chose Copilot to provide a stronger comparison to evaluate \emph{GroundLink}'s in-situ feedforward and spatial mapping features against a robust text-and-video baseline with AI capabilities, rather than creating an artificial comparison with minimal tool support. This allowed us to specifically evaluate the added value of \emph{GroundLink} features. We ran Copilot on a separate Windows laptop to avoid window switching and provide a smoother study experience. The independent variable for this study was the tool difference (treatment: use of \emph{GroundLink}). Participants tried both tools throughout the user study in this within-subjects design.

\autoref{fig:task} shows the study sequence. After obtaining informed consent and providing a brief introduction via presentation slides, we began each session with a task. In each task, participants were assigned a scene (river or forest) and a condition (baseline or \emph{GroundLink}). We then provided a 10-minute hands-on, activity-based walkthrough of the assigned condition using the exercise scene. Next, participants had 20 minutes to complete eight tasks (\autoref{tab:river}, \autoref{tab:forest}). We chose the number of tasks and time budget based on similar experimental setups~\cite{numan2025cocreatar}, adjusted for the unique difficulty of our VP-specific tasks. After 20 minutes, participants completed a survey consisting of the UES-SF~\cite{o2018practical} to measure user engagement, NASA-TLX~\cite{hart1988development} to measure workload, and a post-task questionnaire (\autoref{tab:post-task}) measuring shared mental models (inspired by van Rensburg and colleagues~\cite{van2022five}), transactive memory systems (one question per dimension, modified for this study~\cite{lewis2003measuring}), and perceived conflicts (inspired by Jehn~\cite{jehn1995multimethod}). 
\added{Because our focus was exploratory, and because the experiment happened under a tight timeline}\footnote{This was because participants were high-profile.}\added{, we did not utilize the full, validated (but long to administer) SMM and TMS instruments. Instead, we designed a small subset of customized questionnaire items, adapted from prior work, as a means of quantifying subjective responses related to shared understanding and division of expertise. These items are intended as coarse, indicative measures that complement our qualitative analysis rather than as definitive psychometric evidence. For more conclusive claims about SMM and TMS, future work should employ the complete standardized instruments and larger samples.} We measured engagement and workload to determine whether \emph{GroundLink} impacts task performance. Through the post-task questionnaire, we measured the quality of the common ground that was formed (\textbf{RQ2}), and through the confidence ratings, we measured confidence in task completion~(\textbf{RQ3}).

We repeated this task once more with the scene and condition not yet tried, counterbalancing both scene and condition across participants to control for order effects. After both tasks, participants completed a post-study survey (including preference questions) and a 15-minute semi-structured interview \added{(See \autoref{sect:comparativeQs} for the list of questions)}. The total study took approximately 90 minutes. Prior to the main study, we conducted a pilot session to refine the protocol, which resulted in adjustments to task timing and instruction clarity.

\subsubsection{Measures and Analysis}
We collected survey responses using Likert scales: UES-SF (5-point Likert scale), NASA-TLX (7-point Likert scale), post-task questionnaire (5-point Likert scale, \autoref{tab:post-task}) and task completion and confidence (5-point Likert scale plus ``Did not attempt''). We also collected system usage data (element movements, feature usage), but given the exploratory nature of the study and the creative nature of VP (where there is no strictly correct response), we focused our analysis on survey responses. 

Considering the ordinal nature of these responses, we performed Wilcoxon Signed-Rank tests using scipy 1.7.3~\cite{2020SciPy-NMeth}. We report means for each condition to better illustrate the distribution, particularly given our sample size. 
The choice of our test makes Type I errors less likely when the distribution is possibly skewed or unknown~\cite{bridge1999increasing}, potentially enhancing the rigor of our analysis. To provide a complete picture of results, we supplement significance testing with 95\% bootstrapped confidence intervals (CI) on means, calculated using arch 5.3.1~\cite{kevin_sheppard_2022_6684078} in Python. We report all test statistics and CIs in \autoref{tab:stats_all}. Given the exploratory nature and multiple outcomes, we center interpretation on directions and means rather than significance thresholds. 
We also performed preference testing, with results reported descriptively.

For qualitative analysis, we conducted 15-minute semi-structured interviews at the end of each study session. All interviews were recorded, transcribed, and analyzed using Dovetail\footnote{\url{https://dovetail.com}} following thematic analysis procedures~\cite{braun2006using}, \added{using the same method used in our formative study (\autoref{sect:thematicAnalysis})}.

\subsection{Expert Evaluation}
The goal of the expert evaluation was to evaluate the specific benefits or limitations that \emph{GroundLink} could provide from the perspective of VP professionals.

We recruited 5 experts, including participants from our formative study who had not participated in the comparative study, and additional participants through snowball sampling. Recruited expert participants are denoted as \textbf{E$n$}, all with extensive\footnote{\added{5.6$\pm2.51$ years of experience with 14.0$\pm$9.06 virtual production projects; Roles \\included: Executive Producer, VP Supervisor, VP Artist, Compositor, and \\VP Researcher. All men.}} virtual production experience (demographics: \autoref{tab:interview}). 

The expert evaluation followed a technology probe approach~\cite{park2024coexplorer,hutchinson2003technology} and was conducted remotely via Zoom. Participants gained remote control of the researcher's computer to experience \emph{GroundLink}. Before each session, we manually verified that video streaming latency and control responsiveness were adequate for the evaluation\footnote{For \textbf{E4}, intermittent latency issues were reported during the session.}. 

For each participant, the researcher provided the same study introduction presentation as used in the comparative study. The researcher then provided a tutorial using the exercise scene (identical to the comparative study), gathered first impressions on the features, and asked participants to experience either the river or forest scene (randomly assigned, using the same scenes from the comparative study). Participants completed 2-3 tasks with close researcher guidance\footnote{For \textbf{E5}, researcher demonstration was done instead due to technical difficulties.}.

Each session concluded with a 30-minute in-depth, semi-structured interview \added{(See \autoref{sect:expertQs} for the list of questions)}. All interviews were recorded and transcribed using Dovetail\footnote{\url{https://dovetail.com}}, and we conducted thematic analysis on these recordings~\cite{braun2006using} \added{using the same method used in our formative study (\autoref{sect:thematicAnalysis})}. The study lasted approximately 60 minutes.

\section{Findings}
\begin{figure*}[t]
    \centering
    \includegraphics[width=\linewidth]{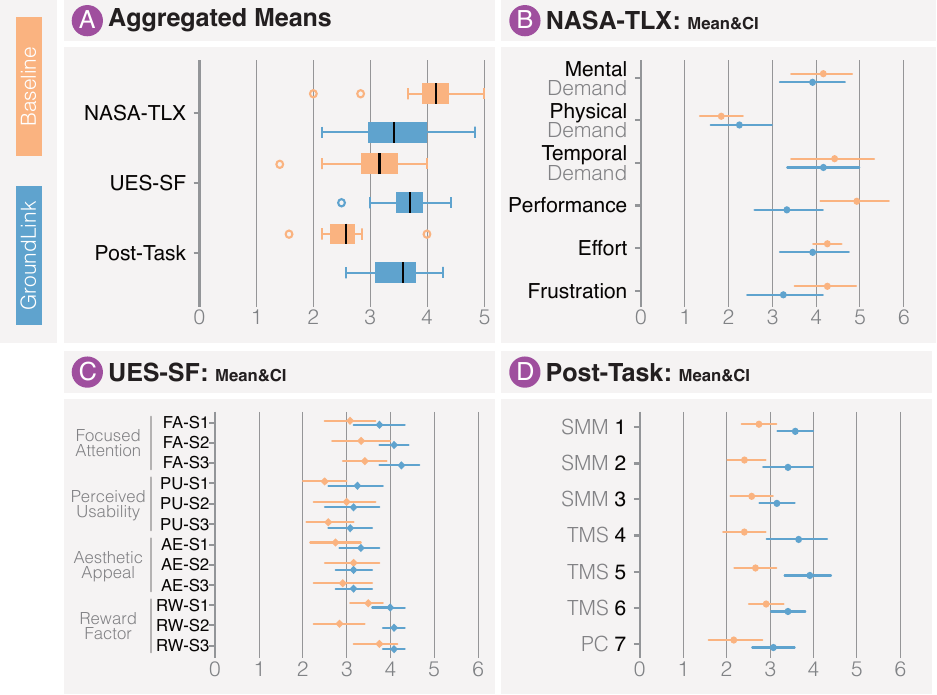}
    \caption{(A) Box plot of aggregated scores for NASA-TLX, UES-SF \added{and Post-Task survey}. (B) Mean and 95\% CI for NASA-TLX (lower indicates lower workload). (C) Mean and 95\% CI for UES-SF. (D) Mean and 95\% CI for the post-task questionnaire (\autoref{tab:post-task}), \protect\rev{which aggregates responses corresponding to SMM, TMS, and perceived conflict-related questions.} Higher values indicate better outcomes for all measures except NASA-TLX, where lower values indicate lower workload (better outcomes). Reverse scoring is used to provide this consistency. SMM: Shared Mental Model, TMS: Transactive Memory System, PC: Perceived Conflict.}
    \Description{The figure has 4 plots. A shows a box plot of aggregated means for the NASA-TLX, UES-SF, and Post-Task survey responses for GroundLink and Baseline. B shows the means and confidence intervals for each dimension of the NASA-TLX across two conditions. C shows per-question means and confidence intervals across two conditions for UES-SF. D shows per-question means and confidence intervals across two conditions for the post-task survey questions.}
    \label{fig:surveys}
\end{figure*}


We found that the comparative study exerted appropriate task load and provided an engaging experience. While both conditions were generally comparable on engagement and task load, we observed some differences. For the UES-SF aggregate (mean; \autoref{fig:surveys}A), \newline\noindent\emph{GroundLink} had a higher mean ($W=17.5, p=0.092$). For NASA-TLX, we did not detect a statistically significant difference ($W=23.0, p=0.23$), though the mean favored \emph{GroundLink} (lower workload). At the item/subscale level, TLX-Performance (reverse-scored) had significantly lower score for \emph{GroundLink} (\autoref{fig:surveys}B; $p=0.042$). For UES-SF, Focused Attention item (FA-S2\textit{``The time I spent using \emph{GroundLink} just slipped away''}) had significantly higher score ($p=0.024$) for \emph{GroundLink}. FA-S3 (\textit{``I was absorbed in this experience.''}) also had higher score for \emph{GroundLink}, but not significant ($p=0.077$). The Reward item RW-S2 (\textit{``My experience was rewarding.''}) had significantly higher score for \emph{GroundLink} ($p=0.011$), with the largest mean difference. These were consistent with CIs (\autoref{fig:surveys}C). 

From this understanding of engagement and task load, it shows that our comparative study has provided appropriate ground for comparing both solutions, showing potentially more focused attention and a higher sense of reward and perceived performance for~\emph{GroundLink}. 

When post-task questionnaire responses were aggregated, it showed a statistically significant difference between the two conditions ($W=8.0, p=0.012$). In the rest of this section, we report how experts and comparative study participants found the experience different, to explore the potential impacts of \emph{GroundLink} in VP scene assembly (editing) task.

\vspace{-0.8em}

\subsection{\emph{GroundLink} \added{Helps Close} the Gaps in Common Ground (RQ2)}
Our \textbf{RQ2} asked: How does surfacing meeting-derived knowledge support a team member in virtual production projects form common ground with existing team members? From the literature research, we found that supporting the formation of Shared Mental Model (SMM) and Transactive Memory System (TMS) would be important in helping users form common ground with the existing team members. \added{Considering the complimentary nature of two theories, we analyzed the results for each separately, and} we found that with \emph{GroundLink}, both experts and comparative study participants began \added{perceiving to form}
 SMM and TMS with the existing team, while foreseeing fewer conflicts with them.

\subsubsection{\added{Participants Perceive Better SMM with \emph{GroundLink}}}
\par\noindent

SMM is an important part of common ground. We found \added{indications to suggest} that with \emph{GroundLink}, both experts and comparative user study participants perceive themselves to be more on the same page with the collaborators in meetings.

\begin{expert}
All experts suggested the benefit of \emph{GroundLink} in this regard, in particular, \textbf{E4}: \textit{``It would be amazing that we are, as a team, on one page. We are landing and standing on one page [with \emph{GroundLink}].''} \textbf{E5} has seen the potential of \emph{GroundLink} in enhancing existing practice of note management in VP: \textit{``Our problem was that after the meeting, someone took notes [...]But again, the context, what people were seeing inside of the volume was missing.''} Experts (\textbf{E1-4}) generally found AI summary helpful in this regard, with particular emphasis on the interaction it provides. As \textbf{E3} summarized: \textit{``You can click these highlighted texts and it just takes you right to that part of the meeting.''}

Experts also found that \emph{GroundLink} could help surfacing important information from past meetings. \textbf{E2,3,5} mentioned the nature of meetings in creative settings, and how \emph{GroundLink} can help collaborators obtain important information without loss, as \textbf{E3} described: \textit{``A lot of things can get lost in translation and things can be misinterpreted or maybe once time goes by, you start to second guess certain things. But when \emph{GroundLink} integrates the meetings, it's really useful for [preventing this].'''} \textbf{E5} emphasized the length of meetings in VP: \textit{``If you just think about a two hour recording and you have multiple people talking. It's very cumbersome to find the right spot.''} 
\end{expert}

\begin{cs}
Similarly, comparative study participants also \added{has shown indications to suggest which they} perceived that they had a better shared mental model with the participants in the meeting recordings with \emph{GroundLink}. For all survey questions relating to SMM, the means were higher for \emph{GroundLink} (\autoref{fig:surveys}D), with significantly higher score for \textit{``I understood why earlier collaborators made the design choices I saw on this screen.''} ($p=0.026$) and \textit{``I was aware of the constraints or conventions that applied when I edited the scene.''} ($p=0.036$).

Comparative study participants \added{perceived that \emph{GroundLink} helped to} obtain the information needed for understanding the team. For \textbf{P10} and \textbf{P11}, the summary view contributed to general understanding of the edits. Similarly, in understanding the mental model of existing team members, \textbf{P4,7-9,11,12} found the way that \emph{GroundLink} aggregated all the changes and comments, and displayed them in their work application helpful. \textbf{P8} suggested: \textit{``I didn't have to go back and forth. I could stay within my working context.''} \textbf{P1,2,8,11,12} found that \emph{GroundLink} helps in understanding more precise implications of past meeting discussions. 
\textbf{P1} mentioned : \textit{``I noticed how much leeway and space there might be [in editing a scene with \emph{GroundLink}]. [...] And I felt like having a visual representation of these ranges made it much easier for me than [having] nothing here, or trying to extract it from text and then resolving these constraints in my head.''} Like experts, comparative study participants also found AI summary useful in general. \textbf{P11} used it for obtaining the overall context, and \textbf{P3,11,12} found the insights from it is precise. Similar to expert, \textbf{P2} also appreciated the idea of surfacing important decisions/comments in work~application.

\end{cs}

\begin{overall}
Our results indicate that proactive provision of applicable information in-situ, at the right time \added{could} have an important role in helping users of the creative software in VP to form a Shared Mental Model, extending how the model could be concretely instantiated with the support of more advanced user interface. 
\end{overall}

\subsubsection{\added{Participants Perceive More Effective TMS with \emph{GroundLink}}}
\par\noindent
\begin{expert}
\textbf{E1,2} particularly appreciated that the interface shows who edited/agreed the particular decisions/comments which can be useful for further discussion on the decision. To support such communication more efficiently, \textbf{E1} suggested that \emph{GroundLink} can be equipped with more advanced share features for each comment/change item: \textit{``I think it would be easier just to have [comments/changes] numbered as well. And then have a share button that would allow you to take those notes, drop them into an email with the context and maybe even a little image that you're sending out to an artist or somebody and say `I need you to take a look at this right now.'''}
\end{expert}

\begin{cs}
The means for all survey questions relating to TMS were higher for \emph{GroundLink}, with significantly higher scores for \textit{``I could easily tell who I would need to consult about a specific asset or decision.''} ($p=0.032$) and \textit{``I trusted that the information I saw about prior decisions was accurate and credible.''} ($p=0.015$) (\autoref{fig:surveys}D). This finding is consistent with the findings from interviews, with \textbf{P3,7,8,10,11} emphasizing the value of showing the editor/agreer on the interface, as \textbf{P7} described: \textit{``Each item has the person assigned to it, and probably this person is the person who made the statement, and that's why this is the person we can reach out to [if we disagree/have~a~question].''}
\end{cs}

\begin{overall}
This result suggests that the important feature to support TMS \replaced{can be}{is} to provide the right relevant profiles of professionals in-situ, on the atomic chunk of information that collaborators consume. 
\end{overall}


\subsubsection{Participants Foresee Fewer Conflicts with the Existing\\ Team Using \emph{GroundLink}}
\par\noindent
\begin{expert}
    \textbf{E1} foresaw reduced conflict with the existing team if \emph{GroundLink} is used in VP: \textit{``I do see a lot of efficiency and a lot more things getting done correctly the first time with [\emph{GroundLink}] because it's right there in front of you. [...] It basically walks you through [the task] and holds your hand.''}
\end{expert}

\begin{cs}
We also observed a score indicating reduced perceived conflict (PC) with \emph{GroundLink} compared to the baseline (``I worried that my changes might conflict with earlier decisions.''\footnotemark; $p=0.062$; \autoref{fig:surveys}D), but not significant.

Half of the participants (\textbf{P5,7-10,12}) worried less about potentially undoing others' work or not following previous decisions with \emph{GroundLink}, as \textbf{P12} suggested: \textit{``On Copilot it's strictly me who has to figure out what to do and then I might be messing something up. But with \emph{GroundLink}, I think I offloaded that part of my brain that would worry about that.''} Also, most participants (\textbf{P1-4,8-12}) thought the scene edited with \emph{GroundLink} would be accepted by the existing team that created the meeting recordings. In contrast, only \textbf{P5,6} thought the scene edited with Copilot would be accepted. \textbf{P9} compared their confidence: \textit{``With \emph{GroundLink}, 75\% versus with Copilot, it's 40\%. Because there is more context with \emph{GroundLink} and then the bounding boxes [exist]. Versus [on Copilot,] you have to piece together.''}
\end{cs}

\begin{overall}
These results suggest that forming a common ground could result in better outcomes in terms of perceived performance and perceived reduction of conflicts.
\end{overall}

\subsubsection{Challenge: Can Make [Junior] Professionals Too Reliant}

\noindent\emph{GroundLink}\footnotetext{Reverse scored} aims to provide powerful features to support forming common ground with the team, which is a realm of communication skill. While the aim is to support, some participants raised concerns about it potentially replacing the skill.

\begin{expert}
\textbf{E1} thought \emph{GroundLink} limits junior professionals from learning how to form common ground: \textit{``If a junior is working on a project with [\emph{GroundLink}] and then they go to another place and they don't have this tool, it's going to be like, `oh, well, I don't know what to do.'''}, but regardless suggested that \emph{GroundLink} could be more agentic, by autonomously generating edits and provide accept/reject options to users. Similarly, \textbf{E2,3} were concerned that \emph{GroundLink}'s feedforward might result in professionals blindly following the system, as \textbf{E2} described: \textit{``I wonder if the person will become too reliant. And then, you'll notice the first thing I did was start ignoring information. And I think a young person would absolutely do that same thing.''}
\end{expert}

\begin{cs}
Consistent with experts, \textbf{P8,9} also suggested that \emph{GroundLink} could perform more agentic edits, which might be inconsistent with the capability of \emph{GroundLink}. \textbf{P9} mentioned: \textit{``Maybe instead of us dragging [elements in scene], \emph{GroundLink} could ask for possible options and then [users] accept it and then see how it looks rather than [users] moving it.''}
\end{cs} 

\begin{overall}
    These perceptions may be due to perceiving \emph{GroundLink} as too powerful. The challenge here seems to be communicating the capabilities of \emph{GroundLink} which remains an open question.
\end{overall}

\subsubsection{Challenge: Authoring Process}
\par\noindent  
\begin{lw}
\noindent In the current implementation of \emph{GroundLink}, researchers manually authored the comments/changes data, but participants were not explicitly told about the authoring process.
\end{lw}

\begin{expert}
\textbf{E3} thought the authoring process could be a potential barrier: \textit{``If it works exactly how it does here [in the study session], that's great. But I'm just suspicious of what it takes for it to get to this.''} Similarly, \textbf{E1,5} suggested that the nature of feedback in video meetings demand a mediator before interpreting the suggestion. \textbf{E5} said: \textit{``In the film industry, when you receive feedback by filmmakers, you might not want to use it exactly word by word [as they might not understand low-level details]. But also sometimes the feedback can be very harsh. So our job is to translate it in a way that makes sense to the artists, without hurting anyone's feelings.''}

\end{expert}

\begin{overall}
Our research focused on exploring the the knowledge consumption process, and as such, authoring process was not part of our scope of exploration. However, future research seems necessary for the authoring process.
\end{overall}

\vspace{0.8em}

\subsection{\emph{GroundLink} Enabled More Confident, \\ Easier VP Scene Editing (RQ3)}

\begin{lw}
\noindent Our RQ3 asked: How does acquiring knowledge from past meeting recordings affect team members' perceived confidence and task difficulty? From both experts and comparative study participants, we found that \emph{GroundLink} might contribute to the increased perceived confidence and reduced task difficulty.
\end{lw}

\begin{expert}
    \textbf{E1-4} found that \emph{GroundLink} can guide the users of the 3D editor. \textbf{E3} found: \textit{``When I [look at feedforward on] the opacity, I can understand why it's like that, and the functionality is there,  [and what action to take].''} \textbf{E4} focused on a more holistic aspect of the feedforward when thinking about it: \textit{``It's a good idea [for a system to] give you an indication of where and to put your [element], especially in terms of visual [cues] and the changes [needed].''} Similarly, \textbf{E3,4} found filtration useful as it maps to contributors and/or scene elements. \textbf{E3} mentioned: \textit{``There are a lot of times when you feel you remember someone mentioning something like the director or the camera person, and you couldn't quite remember what it was. It's nice that you can click on a contributor to see what exactly they wanted.''}
\end{expert}

\begin{figure}[h]
\raggedright
    \includegraphics[width=.95\linewidth]{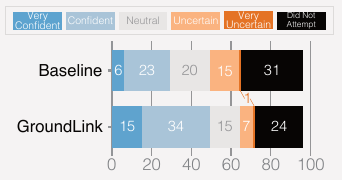}
    \caption{Cumulative confidence ratings across all \\comparative study participants.}
     \label{fig:confidence}
    \Description{Bar charts show the cumulative confidence ratings across participants for the baseline and the GroundLink conditions. From very confident to did not attempt, the baseline counts are 6, 23, 20, 15, 1, and 31. For the GroundLink condition, the counts are 15, 34, 15, 7, 1, and 24.}
\end{figure}
\begin{cs}
Consistently, the number of tasks that comparative study participants confidently attempted\footnotemark was significantly higher (\autoref{fig:confidence}) for \emph{GroundLink} ($p=0.014$). However, the number of tasks attempted between the two different conditions did not significantly differ ($p=0.321$), though the mean favored~\emph{GroundLink}.

This modulation in confidence is also suggested in interviews. \textbf{P1-5,7-10} suggested that they were more confident in their edits when they were using \emph{GroundLink}. \textbf{P3} mentioned: \textit{``[With] \emph{GroundLink} I am more confident. I have some clues or creative messages.''} \textbf{P4} attributed the confidence to the feedforward: \textit{``It gives me [positional constraints] and also a warning message.''} \textbf{P8,9,11,12} also found that \emph{GroundLink} is capable of guiding their edits. \textbf{P8} attributed this to the feedforward mechanism: \textit{``There's more guidance [with \emph{GroundLink}], there's more to the tasks and to the things that need to be done [inferred from the feedforward mechanism].''} For \textbf{P9}, they found the positional feedforward particularly helpful: \textit{``I was moving the light (scene element). There were arrows of where you can move it to. I liked how [\emph{GroundLink}] also highlights the things that are wrong. So [if the] opacity is wrong, it highlights them.''} Similarly, many participants (\textbf{P3,7,8,11,12}) found the mapping between dashboard, meetings, and the 3D editor helpful in referencing scene elements, indicating that the referential links in realizing Dual Coding Theory might be functional in \emph{GroundLink}. For \textbf{P7}, filtration combined with mapping was useful: \textit{``I can now select the objects easily which I can interact with, from this objects filter, which is pretty cool because I can get my task and I can quickly identify this asset in this scene.''}
\end{cs}
\footnotetext{``Confident'' or ``Very Confident'' - Top-2 box}

\begin{overall}
    The results show that providing feedforward and synchronization between 3D editor and dashboard can aid the formation of common ground and increase confidence and ease in 3D editing.
    In doing so, we found a clue in realizing Dual Coding Theory more effectively: the provision of referential link, and visualizing it in-situ on the editor application.
\end{overall}

\subsubsection{Challenge: Visual Clutter and Information Overload}
\par\noindent
\begin{lw}
\noindent\emph{GroundLink} provided more information, directly on the workspace, and inevitably this caused varying levels of visual clutter, confusion, and information overload.
\end{lw}
\begin{expert}
    \textbf{E1,2} found the visual clutter can cause lack of efficiency in editing the 3D scene. \textbf{E1} attributed this to multiple blinking areas: \textit{``Because there are all these green zones flashing right now, but the light (scene element) needed to be moved to a specific spot with all those zones flashing, it could mean that that light can go in any of those [areas].''}

    The amount of information itself was overwhelming to \textbf{E1,2} as well. As part of it, \textbf{E2} attributed this to the extraneous language that AI summary includes: \textit{``I'm not super impressed with the AI summary. It never answered the things I wanted it to answer. [...] The AI responses here are a little flowery.''} Similarly, due to the amount of content provided, it caused some distraction for \textbf{E5}.
\end{expert}

\begin{cs}
Most participants also experienced some difficulties in coping with the visual clutter it creates on the interface (\textbf{P1-5,8,10,11}). \textbf{P4} suggested similarly, but attributed the confusion to the overlaps of effects: \textit{``I think the circles, which overlap, make it confusing. [...] Also, the shape of a circle might give me a sense of uncertainty.''} \textbf{P6,10} found some cues on \emph{GroundLink} confusing because they appeared to demand precision. As \textbf{P6} mentioned: \textit{``I wasn't sure what \emph{GroundLink} wanted although it was being very precise and trying to be precise. I felt frustrated.''} 

Similar to experts, \textbf{P5} found the details added to the AI summary excessive. As \textbf{P5} suggested: \textit{``[AI summary of \emph{GroundLink}] was not useful. It was very hand-wavy without any action. [Copilot] was more direct to the point.''} 

We found the comments on AI summary, though, are more nuanced. We found the exact opposite comments around Copilot's summaries. \textbf{P3,9} found Copilot's text too long (compared to \emph{GroundLink}). \textbf{P3} suggested: \textit{``I found [Copilot's summary] not intuitive enough and not straightforward enough. So I have to read line by line and there are so many lines there.''} For \textbf{P12}, they found Copilot's summary not making sense: \textit{``Honestly, Copilot was just giving me answers that did not make sense. And I felt if I really did follow [Copilot], it would be completely wrong.''}
\end{cs}

\begin{overall}
    The visual clutter, and the amount of information seems to cause varying levels of confusion and overwhelm. But as illustrated in the example of AI summary, it could be due to some preferences of users. Also, \textbf{P4}'s frustration on the uncertainty caused by circular shape of the positional feedforward ironically validates that uncertainty is communicated through \emph{GroundLink}'s design as intended. 
\end{overall}

\subsubsection{Challenge: Perceiving \emph{GroundLink} as Restricting Edits}
\label{sect:challenge_restriction}
\par\noindent
\begin{expert}
    As \textbf{E2} mentioned, participants did not bluntly follow the constraints visualized in \emph{GroundLink}: \textit{``Even if I say as the director, `Oh, the opacity of the water should be 50\%', I'm guessing. [...] And so I was making a creative decision.''} 
\end{expert}

\begin{cs}
Similarly, \textbf{P12} was also exhibiting more conscious edits: \textit{``I did go rogue for some [feedforward effects]. There was one decision specifically where I did not agree with the tool on how to place it. [If these disagreements happen, I would] have talked to my supervisors.''} When users diverged from the restrictions visualized and surfaced, \emph{GroundLink} respected those edits.

However, two participants \textbf{P6,10} found it limited their edits and reduced confidence. As \textbf{P10} suggested: \textit{``I accomplished better, more things in Copilot. Whereas when I was working with \emph{GroundLink}, I felt like I was doing something wrong because there were so many constraints being put. [...] I take more control [with Copilot].''}
\end{cs}

\begin{overall}
    As such, it appears that we would need to find a way to balance the provision of extra information, and reducing the perception of excessive control. We discuss further about this topic in the discussion.
\end{overall}

\subsection{Overall Perception: Participants Preferred \emph{GroundLink}}
\begin{expert}
All experts (\textbf{E1-5}) were excited about the potential impact it may have on virtual production, as \textbf{E1} summarized: \textit{``[GroundLink] allows teams to be a lot more in tune with one another. There's a lot of collaboration [in VP], and it is a lot easier [with \emph{GroundLink}]. Information is right there. There's no second-guessing it because if the information here is wrong, then that's not on the artist or whoever's invoking these changes.''} \textbf{E1-2, E4} particularly emphasized that the concept implemented in \emph{GroundLink} never existed before, as \textbf{E4} suggests: \textit{``[VP] doesn't have a bible application for communicating. [GroundLink] is something huge and something big and I see it has great potential.''} \textbf{E5} mentioned that it might be able to simulate some in-person review:\textit{``You should meet with some of the key people on set, because the moment you are in the volume, you get a different feeling of scale of things. [When] you have [\emph{GroundLink}], you can simulate it quite nicely.''}

Most expert participants (\textbf{E1}, \textbf{E3-5}) also suggested the generalizability of \emph{GroundLink's} design concepts, as \textbf{E1} summarized: \textit{``[\emph{GroundLink's} concept] is something that's awesome for using in Unreal, Maya, or something [which involves] the 3D sense of the world.''}

Experts also suggested the value of \emph{GroundLink} in VP onboarding (\textbf{E1,2,4}) and education at colleges and universities (\textbf{E4}). \textbf{E4} suggested: \textit{``It gives [students] that perception of how they can use and see on their computers and monitors and what's the perfect style.''} 
\end{expert}

\begin{figure}[h]
    \raggedright
    \includegraphics[width=.95\linewidth]{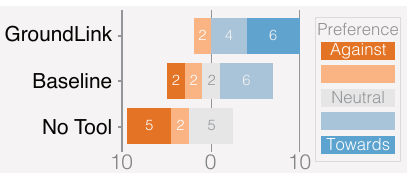}
    \caption{Result of the preference testing.}
    \Description{The figure shows a stacked bar chart of preference testing results. The GroundLink condition has the most preference.}
     \label{fig:preference}
\end{figure}

\begin{cs}
Similar to experts' perception, most comparative study participants preferred working with \emph{GroundLink} when working with meeting recordings for VP. 6/12 comparative study participants strongly preferred working with \emph{GroundLink}, and 4/12 expressed slight preference toward it. This differs from Copilot, which only 6/12 participants slightly preferred working with, and none preferred working without any tool (i.e., just the video player). Only two participants (\textbf{P6, P10}) expressed slight preference against \emph{GroundLink}, likely be due to the perceived restrictive nature of \emph{GroundLink} (\autoref{sect:challenge_restriction}). 
\end{cs}

\section{Discussion}

Our findings reveal how \emph{GroundLink} reshapes common ground formation in virtual production by embedding past meeting decisions directly into the editing workflow. Here, we explore how these results extend collaborative cognition theories, examine the tensions between guidance and autonomy that emerged, and consider broader implications for collaboration systems in virtual production.

\subsection{Forming Common Ground Through \emph{GroundLink}}

Our study reveals a noteworthy finding: Shared Mental Models (SMM) and Transactive Memory Systems (TMS), traditionally viewed as relatively independent cognitive frameworks~\cite{cannon1993shared, wegner1987transactive}, might be simultaneously supported through unified interface design. \emph{GroundLink}'s in-situ feedforward did not just help users know more (SMM) or understand who knows what (TMS). It created the perception that both were simultaneously supported through surfacing contextual information, in-situ.

This integration worked because \emph{GroundLink} provided constituent information when the users needed it. When participants obtained positional constraints in the 3D editor while seeing who made those decisions, they would have built both shared understanding and transactive knowledge. The consistent results across comparative study participants and experts suggest this approach may help address an important challenge in collaborative creative work: the translation gap between discussion and implementation.

This integrative capability of \emph{GroundLink} shows that users no longer need to choose between deep shared knowledge or efficient knowledge distribution. \emph{GroundLink} suggests that appropriate interface design can support both cognitive systems through cohesive information provision.

\vspace{0.5em} 
\noindent\fbox{%
 \parbox{0.47\textwidth}{ 
\textbf{Design Implication 1:} Remote meeting-based collaboration systems can provide constituent information for both shared mental models and transactive memory systems simultaneously, leveraging in-situ presentation, to realize both collaborative cognition systems. 
}}

\subsection{Realizing Dual Coding Theory: The Important Role of Referential Links}

While both \emph{GroundLink} and the baseline provided verbal (meeting-mostly verbal) and non-verbal (3D scene) information separately, our results highlight that simply having multiple representations may not be sufficient. The important difference was the explicit referential linking that \emph{GroundLink} provided between these modalities.

This finding extends Dual Coding Theory~\cite{paivio1991dual} by suggesting that in complex creative tasks, the cognitive challenge is not just processing different information types but maintaining coherence and connection between them. When \textbf{P7} was able to ``quickly identify this asset'' through filtered object selection, they were not accessing new information but leveraging pre-established connections by \emph{GroundLink} that would otherwise require significant mental effort.

The seemingly trivial act of mapping verbal discussions to visual elements may have sizeable impacts on usability, suggesting that VP tools, or even collaborative creative tools more broadly, may need to consider the cognitive burden of maintaining these referential links~manually.
\rev{This also clarifies \emph{GroundLink's} novelty relative to prior VP review/annotation systems, which often keep comments as static marks and/or decouple them from the tool where edits are performed. In contrast, \emph{GroundLink} provides referential links between meeting utterances and scene elements, making meeting-derived decisions actionable in the editor at the moment of manipulation, while preserving provenance back to the originating discussion segment.}

\vspace{0.5em} 
\noindent\fbox{%
\parbox{0.47\textwidth}{ 
\textbf{Design Implication 2:} Systems should prioritize automatic, persistent referential linking between verbal and non-verbal information sources, as the cognitive burden of maintaining these mappings manually may hinder task performance. 
}}

\subsection{Design Friction as a Feature: Balancing Guidance and Autonomy}

The experience of a minority of our participants reveals a fundamental tension in collaborative system design. Most participants found \emph{GroundLink}'s constraints empowering. They felt more confident, possibly because the system provided boundaries. Yet \textbf{P6} and \textbf{P10} experienced these same constraints as restrictive, feeling they had lost control.

This tension became most salient through our atypical participant. \textbf{P6} did not follow the instructions (which were to ``make edits according to what collaborators have discussed in meetings'') despite repeated reminders. Interestingly, this case inadvertently validated our design approach. We posit that \textbf{P6}'s substantial performance difference (3/8 tasks with \emph{GroundLink} vs. 8/8 with baseline) was not a failure but design friction~\cite{cox2016design} working as intended. \textbf{P6}'s admission of being ``confident, but I understood nothing'' with the baseline illustrates the danger of unconstrained editing in collaborative contexts.

However, this friction comes with a drawback: visual clutter. Most participants raised concerns about overwhelming visual feedback. Yet, this same clutter provided the very resistance necessary to prevent thoughtless edits. This suggests that the challenge is not eliminating visual complexity but calibrating it appropriately. \textbf{E1}'s suggestion of (task) priority-based filtering offers a path forward. Interfaces could adapt visual feedback intensity based on the importance of edits rather than displaying all constraints equally.

The fact that \textbf{P12} and \textbf{E2} successfully ``went rogue'' when they disagreed with suggestions demonstrates that \emph{GroundLink} may achieve an important balance: it makes ignoring team decisions require deliberate choice rather than accidental oversight.

\vspace{0.5em} 
\noindent\fbox{%
\parbox{0.47\textwidth}{ 
\textbf{Design Implication 3:} Collaborative editing systems should implement adaptive visual friction that scales with the importance of decisions, ensuring that overriding team decisions demands deliberate intent. However, this should simultaneously minimize unnecessary visual clutter for routine edits. 
}}

\subsection{The Double-Edged Sword of Augmented Collaboration Tool}
\label{sect:doubleEdgedSword}

Expert concerns about junior professionals becoming too reliant on \emph{GroundLink} resemble broader anxieties about the impact of generative AI on critical thinking capabilities~\cite{lee2025impact}. \textbf{E1}'s worry that the system ``holds your hand'' too much suggests a risk of replacing rather than supporting collaborative skills and competencies.

\emph{GroundLink} aims to redistribute rather than eliminate cognitive work. Users are still encouraged to exercise their own judgment, and the interface attempts to support these cognitive processes. The only difference provided by \emph{GroundLink} is that users now simply need to focus on evaluating decisions rather than reconstructing them. The experience of participants overriding the system suggestions when they disagreed indicates that critical thinking remains active.

More intriguing is \textbf{E5}'s observation about feedback translation. In traditional VP workflows, harsh director feedback is diplomatically translated (by coordinators) before reaching artists. \emph{GroundLink}'s mediated presentation might inadvertently solve this emotional labor problem, as the system, by default, removes unnecessary details in the summary view. This may potentially encourage more honest feedback, knowing it will be presented neutrally. Whether this leads to better or worse creative outcomes remains an open question.

The authoring challenge raised by experts points to a deeper issue: whose interpretation becomes the standard? The system currently requires manual authoring, and this process embeds subjective decisions about what constitutes an actionable insight versus mere discussion. This hidden editorial layer could significantly impact team dynamics in ways we do not yet understand. \added{This editorial layer is important because shared understanding is not the same as agreement. \emph{GroundLink} can help users see what their collaborators think, but in its current form it does not help resolve disagreements. 
Here, a suggestion from \textbf{E1} points to one possible direction: equipping each comment or change item with richer sharing features (subsubsection 6.1.2). Future implementations could integrate share features and opinion-aggregation mechanisms}\footnote{For example, reaction buttons or polls.}\added{ so that disagreements are made visible. Such mechanisms could help mitigate overreliance on the system, and mitigate the risk that junior staff simply follow the feedforward~suggestions.}

\added{In addition, future authoring systems could incorporate manual controls with adjustable weights, allowing teams to define shared ``standard'' comments or priorities. Once these are in place, much of the remaining system work becomes a classification problem: matching new data points to similar, previously labeled ones. This is a well-studied problem in machine learning, supporting feasibility~\cite{minaee2021deep,theodoridis2006pattern}. }

\rev{Taken together, these sharing features and adjustable controls could also serve as practical scaffolding for automating our Wizard-of-Oz pipeline. Concretely, a future system could log time-stamped edit events from Unity (object ID, property, old/new values) and align them with time-coded meeting transcripts (and cursor/whiteboard events, when available) using shared timestamps. A lightweight dialog-act classifier could then identify candidate decision/constraint utterances and convert them into structured comment items (e.g., target object, constraint/parameter, rationale, confidence, and source timestamp). Matching becomes an entity-linking and ranking step, which maps utterance to scene objects using object metadata (names/tags), then selects the best match. Finally, per-item sharing and editing can function as a human-in-the-loop confirmation step.}

\added{Designers could also stage cues more carefully. For example, surfacing them only at key moments such as before sign-off, or assign a severity measure to different violations so that only high-severity issues interrupt the creative flow. These strategies offer more concrete ways to balance the creativity-constraint trade-off.}

\vspace{0.5em} 
\noindent\fbox{%
\parbox{0.47\textwidth}{ 
\textbf{Design Implication 4:} Systems mediating collaborative decisions need to make their interpretive/authoring layer transparent, allowing users to understand and question how meeting discussions become actionable constraints while preserving space for professional judgment and creative exploration. 
}}

\subsection{Limitations and Future Work}
\label{sect:futurework}

These findings provide a strong basis for insights and discussion, though some limitations affect their generalizability. Our expert pool's homogeneity (all men, predominantly from our formative study) reflects industry demographics~\cite{forbesGenderGaps} but limits perspective diversity. The overlap with formative participants for the expert evaluation may have created positive bias toward solutions addressing their previously stated pain points. The user study included only a few participants\added{, with short-term exposure}, reducing statistical power\added{, as well as limmiting generalizability}. Given the exploratory nature of the study and the prioritization of avoiding Type II error (hence uncorrected $p$-values), results should be interpreted with caution. \added{Our evaluation relies on qualitative reports and a customized questionnaire rather than validated SMM/TMS instruments. While being appropriate for our exploratory, domain-specific focus, it means our findings provide indicative, qualitative evidence of perceived SMM/TMS formation rather than formal, standardized~measurements.}


Our simplified task design enabled novice participation for the comparative study, but might not capture the full complexity of professional VP workflows involving hours-long meetings, numerous stakeholders, and high-stakes creative decisions. The absence of objective correctness measures, inherent to creative work, makes it difficult to determine whether \emph{GroundLink} actually improves outcomes or merely increases confidence.

The authoring process remains unresolved. This work focused on the consumption experience of meetings, rather than authoring. Extracting actionable decisions from naturalistic meeting discourse for authoring remains unsolved, particularly given the needs for translation between filmmaker vision and technical implementation~(\textbf{E5}). 

Future work should deploy \emph{GroundLink} in authentic virtual production environments over extended periods\added{, with more participants} to understand whether benefits persist or users develop workarounds\added{, as well as for obtaining generalizable insights}. Investigating impacts on upstream communication, such as how knowing feedback will be embedded might change meeting dynamics, could reveal unexpected system effects. \added{Future work could also utilize validated SMM/TMS scales to strengthen the robustness and comparability of the results.} Finally, the risk of homogenizing creative decisions across the industry deserves careful consideration, as these tools shape not just individual edits but collective creative processes.

\added{As \textbf{E1}, \textbf{E3-5} suggested, design concepts considered in this work could generalize to other digital content creation, other domains of the 2D/3D editing}\footnote{E.g., coordinating structural and design changes in architectural BIM models, or propagating design-review decisions into CAD assemblies for manufacturing.}\added{, as well as beyond a workplace tool (e.g., for education). \protect\rev{Our empirical findings suggest the transferable value lies in the mechanism of maintaining referential links between meeting evidence and work objects, so guidance remains traceable to its source. For example, BIM/CAD tools could surface meeting-derived coordination cues linked to the originating discussion, while educational tools could present recorded critique as shareable examples that students can review and revise as their exercises.} While the current version of \emph{GroundLink} focuses on Unity, the same meeting-indexing and in-situ feedforward concepts could also be realized as a family of interoperable add-ons that share a common Electron application for dashboard, allowing constraints and rationales derived from meetings to appear consistently across multiple contents creation tools. Future designers could utilize our generalizable design concepts in other domains, both in professional tool development and HCI research to further extend the reach of our contribution.}
\section{Conclusion}

This paper introduced \emph{GroundLink}, a Unity add-on for virtual production that embeds decisions from prior meetings directly into the 3D editing environment to support the formation of common ground. Across a comparative user study and an expert evaluation, participants reported clearer \added{perceived} shared understanding and awareness of who to consult, greater confidence, and easier scene editing experiences. These findings were also supported by some survey questions. Participants also raised concerns, such as the visual clutter, occasional perceptions of restricted agency, and potential risks of over-reliance. We presented four design implications to inform future work.
\begin{acks}
 We acknowledge the use of Cursor to help us develop the prototype (https://cursor.com/en).

\end{acks}

\bibliographystyle{ACM-Reference-Format}
\bibliography{zotero_synced,references}

\newpage
\onecolumn
\appendix


\section{Expert Interview Participant Details} 
\begin{table}[h] 
\begin{tabular}{@{}llllllllll@{}} \toprule 
\begin{tabular}[c]{@{}l@{}}\textit{PID} \\ (Formative-\\Interview)\end{tabular} & Role & \begin{tabular}[c]{@{}l@{}}Years of \\ VP Experience\end{tabular} & \begin{tabular}[c]{@{}l@{}}Number \\ of Past \\ VP Projects\end{tabular} & \begin{tabular}[c]{@{}l@{}}\textit{PID} \\ (Formative-\\Survey)\end{tabular} & \begin{tabular}[c]{@{}l@{}}\textit{PID} \\ (Expert \\ Evaluation)\end{tabular} & Country & Age & Gender \\ \midrule

\textbf{FI1} & VP Supervisor & 5 & 8 & \textbf{F2} & \textbf{E1} & US& 54&Man \\
\textbf{FI2} & VP Artist & 2 & 3 & \textbf{F10} & \textit{DNP} & South Korea &41 &Man \\
\textbf{FI3} & VP Artist & 5 & 30 & \textbf{F15} & \textbf{E4} & Canada& 35& Man\\
\textbf{FI4} & Compositor & 10 & 12 & \textbf{F17} & \textbf{E3} & Canada& 28& Man\\
\textbf{FI5} & VP Lead/Researcher & 8 & 15 & \textbf{F22} &\textit{DNP} & Canada& 38& Man\\
\textbf{FI6} & VP Supervisor & 5 & 4 & \textbf{F5} & \textit{DNP} & Canada& 46&Man \\
\textit{DNP} & VP Researcher & 4 & 10 & \textbf{F23} & \textbf{E2}& Canada& 40& Man\\
\textit{DNP} & Executive Producer & 4 & 10 & \textit{DNP} & \textbf{E5}& US& \textit{DNR}&Man \\
\bottomrule 
\end{tabular} 
\caption{Participant information for the semi-structured, expert interviews. \textit{PID}: Participant ID, \textit{DNP}: Did Not Participate, \textit{DNR}: Did Not Report.} 
\label{tab:interview} 
\end{table}

\section{\added{Formative Study - Survey}}
\label{sect:formativeSQs}

\begin{table}[h]
\begin{tabular}{@{}lll@{}}
\toprule
\rowcolor[HTML]{656565} 
{\color[HTML]{FFFFFF} \textbf{ID}} & {\color[HTML]{FFFFFF} \textbf{Question}}                                                                                                                                                                                                               & {\color[HTML]{FFFFFF} \textbf{Response Type}}                                                \\ \midrule
1                                  & Which parts of the virtual production workflow are you mostly involved in?                                                                                                                                                                             & Text                                                                                         \\
\rowcolor[HTML]{C0C0C0} 
2                                  & What is your experience with virtual production workflows?                                                                                                                                                                                             & Text                                                                                         \\
3                                  & \begin{tabular}[c]{@{}l@{}}What are the main steps in your virtual production workflow?\\ (Is it more waterfall-like or iterative process between phases? \\ What tools do you use in each step? \\ Were those Remote/In-person/On-site?)\end{tabular} & Text                                                                                         \\
\rowcolor[HTML]{C0C0C0} 
4                                  & \begin{tabular}[c]{@{}l@{}}How often do you collaborate with others? (For each group)\\ - Staff from different roles\\ - Staff in the same role\end{tabular}                                                                                           & \begin{tabular}[c]{@{}l@{}}5-Point Likert\\ (Never-Always)\end{tabular}                      \\
5                                  & What are the most common issues you encounter in virtual production?                                                                                                                                                                                   & Text                                                                                         \\
\rowcolor[HTML]{C0C0C0} 
6                                  & \begin{tabular}[c]{@{}l@{}}How do you record and resolve these issues? \\ In terms of: Frequency; Recording method (e.g., Jira, Word document); \\ Resolution method (e.g., Zoom meeting, VR meeting, email).\end{tabular}                             & Text                                                                                         \\
7                                  & Do you use Augmented Reality (AR) systems in your virtual production?                                                                                                                                                                                  & Yes/No                                                                                       \\
\rowcolor[HTML]{C0C0C0} 
8                                  & When do you use AR systems and why?                                                                                                                                                                                                                    & Text                                                                                         \\
9                                  & \begin{tabular}[c]{@{}l@{}}Do you review 3D scenes using AR technology? \\ Describe its usefulness or problems.\end{tabular}                                                                                                                           & Text                                                                                         \\
\rowcolor[HTML]{C0C0C0} 
10                                 & Have you onboarded staff(s) to a 3D scene created by your team?                                                                                                                                                                                        & Yes/No                                                                                       \\
11                                 & \begin{tabular}[c]{@{}l@{}}{[}If yes{]} Describe how you onboarded others to a scene. \\ (e.g., meeting format, devices used, software used for onboarding)\end{tabular}                                                                               & Text                                                                                         \\
\rowcolor[HTML]{C0C0C0} 
12                                 & Have you been onboarded to a 3D scene?                                                                                                                                                                                                                 & Yes/No                                                                                       \\
13                                 & \begin{tabular}[c]{@{}l@{}}{[}If yes{]} Describe how you were onboarded to a scene.\\ (e.g., meeting format, devices used, software used for onboarding)\end{tabular}                                                                                  & Text                                                                                         \\
\rowcolor[HTML]{C0C0C0} 
14                                 & How often do onboarding sessions to a scene occur?                                                                                                                                                                                                     & \begin{tabular}[c]{@{}l@{}}7-Point Likert\\ (Daily-Never)+Other\end{tabular}                 \\
15                                 & \begin{tabular}[c]{@{}l@{}}Rate the effectiveness of existing onboarding methods as: (For each case)\\ - A person being onboarded\\ - A person onboarding others\end{tabular}                                                                          & \begin{tabular}[c]{@{}l@{}}5-Point Likert\\ (Very Problematic-\\ Very Good)+N/A\end{tabular} \\ \bottomrule
\end{tabular}
\caption{\added{Survey questionnaire used for the formative study.}}
\end{table}

\newpage

\section{\added{Formative Study - Semi-Structured Interview Questions}}
\label{sect:formativeQs}
\added{Please think about your virtual production experiences in general.}

\begin{itemize}
    \item \added{How would you describe your virtual production experiences? (Compared to other production practices?)}

    \item \added{Do you collaborate with others? (Frequency, context; follow-up from survey responses)}
    \begin{itemize}
        \item \added{How do you collaborate (tools, methods, etc.)?}
        \item \added{Are there any challenges in existing collaborative practices? (Follow-up from the survey)}
        \item \added{Were there any instances when you were missing information from other collaborators?}
        \begin{itemize}
            \item \added{How did you resolve this issue?}
        \end{itemize}
        \item \added{Were there any instances when you were not sure about the intents of other collaborators?}
        \begin{itemize}
            \item \added{How did you resolve this issue?}
        \end{itemize}
        \item \added{Were there instances where you had to redo some of the work others had done, or others modified your work without knowing your intent?}
        \begin{itemize}
            \item \added{How did you resolve this issue?}
        \end{itemize}
    \end{itemize}

    \item \added{Do you review past meeting videos or issue notes (e.g., from version control or production management tools like Flow)?}
    \begin{itemize}
        \item \added{How do you review the notes or videos?}
        \item \added{What works well with these? Are there any challenges?}
    \end{itemize}
\end{itemize}

\added{Please think about an instance when you onboarded a staff member to a 3D scene you have been working on.}

\begin{itemize}
    \item \added{Could you walk me through how you introduced the 3D scene?}
    \begin{itemize}
        \item \added{Which part of the overall workflow was the project in when you introduced the new scene to the other person?}
        \item \added{Were there any constraints you established during the creation of the 3D scene (e.g., object should not be placed here, lighting should be this way)?}
        \begin{itemize}
            \item \added{(If yes) What were those constraints?}
            \item \added{(If yes) How problematic were those constraints if violated?}
            \item \added{(If yes) How did you communicate those constraints?}
            \item \added{(If yes) How frequently, and how substantially, did collaborators violate these constraints?}
        \end{itemize}
        \item \added{Are there other types of information you try to communicate to collaborators?}
        \item \added{How easy or difficult is it to communicate that information? Why?}
        \begin{itemize}
            \item \added{How long does a new collaborator usually take to get used to the scene?}
            \item \added{Are those new collaborators in the same role as you, or not?}
            \item \added{Do those scenes require frequent collaboration? Cross-functional collaboration?}
        \end{itemize}
    \end{itemize}
\end{itemize}

\added{Now, let’s think about the opposite scenario.}

\begin{itemize}
    \item \added{Do you get introduced to a new 3D scene as part of your job?}
    \begin{itemize}
        \item \added{(If yes) How did other collaborators introduce the scene to you?}
        \begin{itemize}
            \item \added{When was it? What was the situation like?}
            \item \added{Any success stories? Or any instance of an introduction that did not work well?}
            \item \added{When you were introduced to a scene, were there instances when you performed well or poorly while collaborating with others?}
        \end{itemize}
        \item \added{Do you think there could have been aspects that, if implemented, would have helped you get used to the scene faster?}
    \end{itemize}
\end{itemize}

\section{\added{Comparative Study - Semi-Structured Interview Questions}}
\label{sect:comparativeQs}
\begin{itemize}
    \item \added{Describe one change you made in the scene and what information led you to make that change.}

    \item \added{Did you ever worry that you might undo someone else's work or might not consider the decisions made by others? What helped to reassure you?}

    \item \added{How confident are you that your final edits will be accepted? Why?}

    \item \added{Were there any points that were the most helpful?}

    \item \added{Were there any points that were the most confusing or unhelpful?}

    \item \added{Do you think accessing the context from different projects might be useful?}
    \begin{itemize}
        \item \added{Under what circumstances?}
        \item \added{What kind of information?}
    \end{itemize}

    \item \added{Any suggestions for improvements?}
\end{itemize}

\newpage

\section{\added{Expert Evaluation - Semi-Structured Interview Questions}}
\label{sect:expertQs}
\begin{itemize}
    \item \added{In what ways do you think the effects happening on the Unity editor or the inspector help or deter your work?}
    \begin{itemize}
        \item \added{Why?}
        \item \added{What are the opportunities or challenges you could foresee from it?}
    \end{itemize}

    \item \added{There was a separate panel for providing a summarized meeting detail. In what ways do you think it would help or hinder the completion of your edits?}
    \begin{itemize}
        \item \added{When you interact with one panel, how would you interact with the other panel?}
        \begin{itemize}
            \item \added{How would the system help or hinder you from performing those tasks?}
        \end{itemize}
    \end{itemize}

    \item \added{There was a filter tool with an AI summary. In what ways would you use filtration and AI summary?}
    \begin{itemize}
        \item \added{In what ways do you think this will be helpful or unhelpful?}
    \end{itemize}

    \item \added{What do you think are the strengths and weaknesses of the tool?}

    \item \added{We are envisioning using the tool for more junior virtual production professionals. In what ways do you think the tool can help or hinder their completion of the task?}
    \begin{itemize}
        \item \added{In what ways do you think they will know or not know the knowledge established within the team?}
        \item \added{In what ways do you think they will have confidence in the edits that they performed?}
        \item \added{How would the usage of the tool help or hinder the user in figuring out who to contact if they have questions about previous edits being done?}
    \end{itemize}

    \item \added{Do you have any suggestions for improvements?}
\end{itemize}
\vspace{-3mm}

\section{Questionnaires}


\begin{table}[H]
\begin{tabular}{@{}lll@{}}
\toprule
ID & Relates To & Question                                                                              \\ \midrule
1  & SMM        & I understood why earlier collaborators made the design choices I saw on this screen.  \\
2  & SMM        & I was aware of the constraints or conventions that applied when I edited the scene.   \\
3  & SMM        & While working, I felt ``on the same page'' with the existing team.                    \\
4  & TMS        & I could easily tell who I would need to consult about a specific asset or decision.   \\
5  & TMS        & I trusted that the information I saw about prior decisions was accurate and credible. \\
6  & TMS        & My edits fit smoothly with what the team had already done.                            \\
7  & PC         & I worried that my changes might conflict with earlier decisions.$^*$                      \\ \bottomrule
\end{tabular}
\caption{Post-task questionnaire used to quantify participants' subjective, qualitative experiences relating to theories. \textit{SMM}: Shared Mental Model, \textit{TMS}: Transactive Memory System, \textit{PC}: Perceived level of conflict. $^*$: reverse scored.} 
\label{tab:post-task} 
\end{table}

\begin{table}[H]
\begin{tabular}{@{}ll@{}}
\toprule
ID & Task Description                                                                               \\ \midrule
1  & Boat needs to be moved and edited to be appearing like an illusion.                            \\
2  & Tent needs to be relocated.                                                                    \\
3  & Rock should be moved: it is not usable by the actors.                                          \\
4  & The sub camera needs to show the boat coming behind Alice and Bob.                             \\
5  & A log should be floating on the river, and more food items need to be placed around the scene. \\
6  & The light source should appear more low-angled.                                                \\
7  & The paddle should be relocated.                                                                \\
8  & The main camera needs to capture wider area, while being located closer to the actors.         \\ \bottomrule
\end{tabular}
\caption{List of tasks for the river scene.} 
\label{tab:river} 
\end{table}

\begin{table}[H]
\begin{tabular}{@{}ll@{}}
\toprule
ID & Task Description                                                               \\ \midrule
1  & Relocate and adjust the mushroom table.                                        \\
2  & Adjust the sub camera. It is currently a placeholder only.                     \\
3  & Adjust the main directional light.                                             \\
4  & Move the rock: left of the tent, and color should follow the style guidelines. \\
5  & Relocate the view-blocking tree (Spruce 6).                                    \\
6  & Bring the campfire closer to the tent (entrance).                              \\
7  & Place the log around campfire.                                                 \\
8  & Place decorative mushrooms in the scene (use Mushroom Large).                  \\ \bottomrule
\end{tabular}
\caption{List of tasks for the forest scene.} 
\label{tab:forest} 
\end{table}

\begin{table}[H]
\begin{tabular}{@{}llllllll@{}}
\toprule
\textbf{\begin{tabular}[c]{@{}l@{}}Questionnaire/\\ Measure\end{tabular}} & \textbf{ID}                                                                & \textbf{$W$} & \textbf{$p$}   & \textbf{\begin{tabular}[c]{@{}l@{}}Baseline \\ Mean\end{tabular}} & \textbf{\begin{tabular}[c]{@{}l@{}}Baseline \\ 95\% CI\end{tabular}} & \textbf{\begin{tabular}[c]{@{}l@{}}\emph{GroundLink}\\ Mean\end{tabular}} & \textbf{\begin{tabular}[c]{@{}l@{}}\emph{GroundLink}\\ 95\% CI\end{tabular}} \\ \midrule
\textbf{NASA-TLX}                                                         & Mental Demand                                                              & 17.0         & 0.507          & 4.17                                                              & {[}3.42, 4.83{]}                                                     & 3.92                                                                                       & {[}3.17, 4.67{]}                                                                              \\
(Lower: Better)                                                           & Physical Demand                                                            & 19.0         & 0.374          & 1.83                                                              & {[}1.33,2.33{]}                                                      & 2.25                                                                                       & {[}1.58, 3.00{]}                                                                              \\
                                                                          & Temporal Demand                                                            & 34.5         & 0.733          & 4.42                                                              & {[}3.42, 5.33{]}                                                     & 4.17                                                                                       & {[}3.33, 5.00{]}                                                                              \\
                                                                          & Performance                                                                & 13.0         & 0.042 & 4.92                                                              & {[}3.42, 5.33{]}                                                     & 3.33                                                                                       & {[}2.58, 4.17{]}                                                                              \\
                                                                          & Effort                                                                     & 30.0         & 0.519          & 4.25                                                              & {[}3.92, 4.58{]}                                                     & 3.92                                                                                       & {[}3.17, 4.75{]}                                                                              \\
                                                                          & Frustration                                                                & 11.0         & 0.168          & 4.25                                                              & {[}3.50, 4.92{]}                                                     & 3.25                                                                                       & {[}2.42, 4.17{]}                                                                              \\
\textbf{UES-SF}                                                           & FA-S1                                                                        & 12.0         & 0.209          & 3.08                                                              & {[}2.50, 3.67{]}                                                     & 3.75                                                                                       & {[}3.17, 4.33{]}                                                                              \\
(Higher: Better)                                                          & FA-S2                                                                        & 0.0          & 0.024 & 3.33                                                              & {[}2.67, 4.00{]}                                                     & 4.08                                                                                       & {[}3.75, 4.42{]}                                                                              \\
                                                                          & FA-S3                                                                        & 10.5         & 0.077 & 3.42                                                              & {[}2.92, 3.92{]}                                                     & 4.25                                                                                       & {[}3.75, 4.67{]}                                                                              \\
                                                                          & PU-S1                                                                        & 13.0         & 0.133          & 2.50                                                              & {[}2.00, 3.00{]}                                                     & 3.25                                                                                       & {[}2.58, 3.83{]}                                                                              \\
                                                                          & PU-S2                                                                        & 24.0         & 0.718          & 3.00                                                              & {[}2.25, 3.67{]}                                                     & 3.17                                                                                       & {[}2.50, 3.75{]}                                                                              \\
                                                                          & PU-S3                                                                        & 16.0         & 0.232          & 2.58                                                              & {[}2.08, 3.17{]}                                                     & 3.08                                                                                       & {[}2.58, 3.58{]}                                                                              \\
                                                                          & AE-S1                                                                        & 5.5          & 0.140          & 2.75                                                              & {[}2.17, 3.33{]}                                                     & 3.33                                                                                       & {[}2.83, 3.75{]}                                                                              \\
                                                                          & AE-S2                                                                        & 17.0         & 0.886          & 3.17                                                              & {[}2.50, 3.75{]}                                                     & 3.17                                                                                       & {[}2.75, 3.58{]}                                                                              \\
                                                                          & AE-S3                                                                        & 28.0         & 0.642          & 2.92                                                              & {[}2.25, 3.58{]}                                                     & 3.17                                                                                       & {[}2.75, 3.58{]}                                                                              \\
                                                                          & RW-S1                                                                        & 6.0          & 0.165          & 3.50                                                              & {[}3.08, 3.83{]}                                                     & 4.00                                                                                       & {[}3.58, 4.33{]}                                                                              \\
                                                                          & RW-S2                                                                        & 0.0          & 0.011 & 2.83                                                              & {[}2.25, 3.42{]}                                                     & 4.08                                                                                       & {[}3.83, 4.33{]}                                                                              \\
                                                                          & RW-S3                                                                        & 6.0          & 0.317          & 3.75                                                              & {[}3.17, 4.17{]}                                                     & 4.08                                                                                       & {[}3.83, 4.33{]}                                                                              \\
\textbf{Post-task}                                                        & 1                                                                          & 9.0          & 0.026 & 2.75                                                              & {[}2.33, 3.17{]}                                                     & 3.58                                                                                       & {[}3.17, 4.00{]}                                                                              \\
\textit{(\autoref{tab:post-task})}                                                          & 2                                                                          & 5.0          & 0.036 & 2.42                                                              & {[}2.00, 2.92{]}                                                     & 3.42                                                                                       & {[}2.83, 4.00{]}                                                                \\
       (Higher: Better)                                                                   & 3                                                                          & 15.0         & 0.191          & 2.58                                                              & {[}2.08, 3.08{]}                                                     & 3.17                                                                                       & {[}2.75, 3.58{]}                                                                              \\
                                                                          & 4                                                                          & 7.0          & 0.032 & 2.42                                                              & {[}1.92, 2.92{]}                                                     & 3.67                                                                                       & {[}2.92, 4.33{]}                                                                              \\
                                                                          & 5                                                                          & 4.0          & 0.015 & 2.67                                                              & {[}2.17, 3.17{]}                                                     & 3.92                                                                                       & {[}3.33, 4.42{]}                                                                              \\
                                                                          & 6                                                                          & 14.0         & 0.156          & 2.92                                                              & {[}2.50, 3.33{]}                                                     & 3.42                                                                                       & {[}3.00, 3.83{]}                                                                              \\
                                                                          & 7                                                                          & 5.0          & 0.062 & 2.17                                                              & {[}1.58, 2.83{]}                                                     & 3.08                                                                                       & {[}2.58, 3.58{]}                                                                              \\
\textbf{\begin{tabular}[c]{@{}l@{}}Post-task \\ Confidence\end{tabular}}  & \begin{tabular}[c]{@{}l@{}}Number of Tasks\\ Completed\end{tabular}        & 18.0         & 0.321          & 5.42                                                              & {[}4.75, 6.17{]}                                                     & 6.00                                                                                       & {[}5.17, 6.83{]}                                                                              \\
(Higher: Better)                                                          & \begin{tabular}[c]{@{}l@{}}Perceived Confidence\\ (Top-2 box)\end{tabular} & 2.0          & 0.014 & 2.42                                                              & {[}1.58, 3.33{]}                                                     & 4.08                                                                                       & {[}3.00, 5.00{]}                                                                              \\ \bottomrule
\end{tabular}
\caption{List of all the test statistics and confidence intervals. 
} 
\label{tab:stats_all} 
\end{table}


\end{document}